\documentclass[preprint,authoryear,11pt]{elsarticle}

\usepackage{amssymb,amsmath}

\usepackage[utf8]{inputenc}
\usepackage{graphics,graphicx}
\usepackage{color}
\usepackage[top=1.in, bottom=1.in, left=1.in, right=1.in]{geometry}

\def\RR{{\rm I\kern-.17em R}}
\def\ZZ{{\rm Z\kern-.32em Z}}
\def\NN{{\rm I\kern-.20em N}}
\def\e{\varepsilon}

\journal{Mechanics Research Communications}

\begin{document}
\begin{frontmatter}

\title{An asymptotic approach to the adhesion of thin stiff films}


\author[label1,label2]{S. Dumont}
\author[label2]{F. Lebon}
\author[label3]{R. Rizzoni}

\address[label1]{Laboratoire Ami\'enois de Math\'ematique Fondamentale et Appliqu\'ee, CNRS UMR 7352, UFR des Sciences, 33, rue Saint-Leu, 80039 Amiens Cedex 1, France, dumont@lma.cnrs-mrs.fr}
\address[label2]{Laboratoire de M\'ecanique et d'Acoustique, CNRS UPR 7051, Centrale Marseille, Universit\'e Aix-Marseille, 31 Chemin Joseph Aiguier, 13402 Marseille Cedex 20, France, lebon@lma.cnrs-mrs.fr}
 \address[label3]{Dipartimento di Ingegneria, Universit\`a di Ferrara, Via Saragat 1, 44122 Ferrara, Italy, rizzoni.raffaella@unife.it}

\begin{abstract}
In this paper, the asymptotic first order analysis, both mathematical and numerical, of two structures bonded together is presented. Two cases are considered, the gluing of an elastic structure with a rigid body and the gluing of two elastic structures. The glue is supposed to be elastic and to have its stiffness of the same order than those of the elastic structures.
An original numerical method is developed to solve the mechanical problem of stiff interface at order 1, based on the Nitsche's method.
Several numerical examples are provided to show the efficiency of both the analytical approximation and the numerical method.
\end{abstract}


\end{frontmatter}

\section{Introduction}
Adhesive bonding is an assembly technique often used in structural mechanics.
In bonded composite structures, the thickness of the glue is much smaller than the other  dimensions.
For example,
in  \cite{GRD08}, the thickness of the glue is equal to 0.1 mm, whereas the dimension of the structure is close to 150 mm, thus the ratio of dimensions of the  bodies considered is
close to $\frac{1}{1500}$. Thus, the thickness of the glue can be considered as a small parameter in the modeling process. Usually, the stiffness of the glue is considered to be one another small parameter
when compared with the stiffness of the adherents (soft interface theory), as shown in \cite{LRR04, LR08}. For example, in \cite{GRD08}, two steel structures are bonded by a Loctite 300 glue and the ratio between the Young moduli of the both components is close to $\frac{1}{230}$.
Nevertheless, in the case of an epoxy based  adhesive bonding of two aluminium structures, the ratio between the Young moduli  is typically about $\frac{1}{20}$(see for example  \cite{CBCS11}). Thus, the stiffness of the glue can not be considered as the smallest parameter (stiff interface theory). The aim of this paper is to analyze mathematically and numerically the asymptotic behavior of bonded structures in the case of only one small parameter: the thickness. In the following, the stiffness is not a small parameter, that is to say the Young moduli of the glue and of the adherents are of the same order of magnitude.

The mechanical behavior of thin films between elastic adherents was studied by several authors: \cite{ACM98,B2006, bigo07a,bigo07b,bigo02,cognard06, CBCS11,duong, GRD08,hirschberger,KL00,kumar10, kumarM11, LRR04,LR08,LR2010,LR2011,LR2011b,LR2007,nguyen,LR2012,elioefred}.
The analysis was based on the classic idea that a very thin adhesive film can be replaced by a contact law, like in \cite{ACM98}. The contact law describes the asymptotic behavior of the film in the limit as its thickness goes to zero and it prescribes the jumps in the displacement (or in the displacement rate) and in the traction vector fields at the limit interface. The formulation of the limit problem involves the mechanical and the geometrical properties of the adhesive and the adherents, and in \cite{LRR04,LR08,LR2010,LR2011,LR2011b,LR2012,LR2007} several cases were considered: soft films (\cite{K91,LRR04}); adhesive films governed by a non convex energy (\cite{OLLM96, LR08,LM97}; imperfect gluing \cite{Z2000}); flat linear elastic films having stiffness comparable with that of the adherents and giving rise to imperfect adhesion between the films and the adherents (\cite{LR2010,LR2011}); joints with mismatch strain between the adhesive and the adherents, see for example \cite{LR2012}. Several mathematical techniques can be used to perform the asymptotic analysis:
 $\Gamma$-convergence,
 Variational analysis,
 Matched asymptotic expansions
and Numerical studies (see \cite{LR2011b,SP80} and references therein).
\medskip



The first part of the paper is devoted to extend the imperfect interface law  given in \cite{LR2011} to the case of  a very thin interphase whose stiffness  is of the same order of magnitude as that of the adherents, firstly when an elastic body is glued to a rigid base, and secondly in the plane strain case.

In the second part of the paper, numerical methods adapted to solve the limit problems obtained in the first part are developed. In the case of the gluing of a deformable body with a rigid solid, the numerical scheme is very classical. On the contrary, the gluing of two deformable bodies leads to more complicated numerical strategies. The proposed method is based on an original method presented in \cite{N74}. This kind of method is well known in the domain decomposition context. This method is implemented in a finite element software.

In the third part, some numerical examples are presented and the numerical results are analyzed (in terms of mechanical interpretation, computed time, convergence, etc.) in order to quantify and justify the methodology. 


\section{Theoretical results for thin stiff films}

\subsection{Asymptotic analysis for an elastic body glued to a rigid base}

Let us consider a linear elastic body $\Omega\subset \RR^3$ of boundary $\partial \Omega$.
This structure is divided into three parts (see figure \ref{fig:app}): two parts (the adherents) are perfectly bonded with a very thin third one (the interphase).
One of the two adherents is considered as rigid. The glue is perfectly bonded with the rigid body.

More precisely, after introducing a small parameter $\e>0$ which is the thickness of the glue, we define the following domains
\begin{itemize}
\item $B^\e=\{ (x_1,x_2, x_3)\in \Omega:\ 0<x_3<\e  \}$ (the glue);

\item $\Omega^\e_+ =\{ (x_1,x_2,x_3)\in \Omega:\  x_3>\e  \}$ (the deformable adherent);

\item $S^\e_+=\{ (x_1,x_2,x_3)\in \Omega:\ x_3=\e  \}$;

\item $\Gamma=\{ (x_1,x_2,x_3)\in \Omega:\ x_3=0  \}$ (the interface);

\item $\Omega_+ =\{ (x_1,x_2,x_3)\in \Omega:\  x_3>1  \}$;

\item $B=\{ (x_1,x_2, x_3)\in \Omega:\ 0<x_3<1  \}$;

\item $S_+=\{ (x_1,x_2,x_3)\in \Omega:\ x_3=1  \}$;

\item $\Omega^0_+ =\{ (x_1,x_2,x_3)\in \Omega:\  x_3>0  \}$.
\end{itemize}

On a part $\Gamma_1$ of $\partial \Omega$, an external load $g$ is applied, and on a part $\Gamma_0$ of $\partial \Omega$
such that $\Gamma_0\cap \Gamma_1=\emptyset$
a displacement $u_d$ is imposed. Moreover, we suppose that $\Gamma_0\cap B^\e = \emptyset$ and
 $\Gamma_1\cap B^\e = \emptyset$. A body force $f$ is applied in $\Omega^\e_+$.

 We consider also that the interface $\Gamma$ is a plane normal to the third direction $e_3$.
 \bigskip


We are interested in the equilibrium of such a structure. The equations of the problem are written as follows:

\begin{equation}\label{eqeq}
\left\{\begin{array}{ll}
\mbox{div} \sigma^\e +{ f}=0 \qquad &\mbox{in }\Omega^\e_+\cup B^\e \\
\sigma^\e n=g & \mbox{on } \Gamma_1 \\
u^\e=u_d& \mbox{on } \Gamma_0 \\
u^\e=0 &\mbox{on } \Gamma \\
\sigma^\e=A_+ e(u^\e) &\mbox{in }\Omega^\e_+ \\
\sigma^\e=\hat A e(u^\e) &\mbox{in } B^\e \\
\end{array}\right.
\end{equation}

where $e(u^\e)$ is the strain tensor ($e_{ij}(u^\e)=\frac12(u_{i,j}+u_{j,i})$, $i,j=1,2, 3$) and
$A_+$, $\hat A$ are the elasticity tensors of the deformable adherent and the adhesive, respectively.
In the sequel, we consider that the glue is isotropic, with Lam\'e's coefficients equal
to $\hat \lambda$ and $\hat \mu$  in the
interphase $B^\e$.
Let us emphasize that the Lamé's coefficients of the interphase do not depend on the thickness
$\e$ of the interphase (this will be referred as the case of a stiff interface hereinafter).
\bigskip

Since the thickness of the interphase is very small, it is natural to seek the solution of problem
(\ref{eqeq}) using asymptotic expansions with respect to the parameter $\e$:
\begin{equation}\label{mae}
\left\{\begin{array}{l}
u^\e=u^0+\e u^1+o(\e)\\
\sigma^\e=\sigma^0+\e \sigma^1+o(\e)\\
\end{array}\right.
\end{equation}


In order to write the equations verified by $u^0$, $u^1$, $\sigma^0$ , $\sigma^1$
in $\Omega^+$ and on the interface $\Gamma$, we consider the method developed
by \cite{LR2011} and based on the mechanical energy of the system:

\begin{equation}
J^\e(u^\varepsilon)=\frac12 \int_{\Omega^\e_+} A_+ e( u^\e)\cdot e( u^\e) dx -\int_{\Gamma_1}g u\ ds +
\frac12 \int_{B^\e} {\hat A} e( u^\e)\cdot e(u^\e)\ dx
\end{equation}

 which is defined in the set of displacements
 \begin{equation}
 V^\e=\left\{ u\in H(\Omega;\RR^3):\ u=u_d\mbox{ on } \Gamma_0, \ u=0  \mbox{ on } \Gamma \right\},
 \end{equation}
 where $H(\Omega ;R^3)$  is the set of admissible displacements defined on $\Omega$.

At this level, the domain is rescaled using the classical procedure:
\begin{itemize}
\item In the glue, we define the following change of variable
$$
(x_1,x_2,x_3)\in B^\e\rightarrow (z_1,z_2,z_3) \in B, \mbox{ with } (z_1,z_2,z_3)=(x_1,x_2,\frac{x_3}{\e})
$$
and we denote $\hat u^\e(z_1,z_2,z_3)=u^\e(x_1,x_2,x_3)$.

\item In the adherent, we define the following change of variable
$$
(x_1,x_2,x_3)\in \Omega^\e_+\rightarrow (z_1,z_2,z_3) \in \Omega_+, \mbox{ with } (z_1,z_2,z_3)=(x_1,x_2,x_3+1-\e)
$$
and we denote  $\bar u^\e(z_1,z_2,z_3)=u^\e(x_1,x_2,x_3)$. We suppose that the external forces and the prescribed displacement
$u_d$ are assumed to be independent of $\e$. As a consequence, we define $\bar f(z_1,z_2,z_3)=f(x_1,x_2,x_3)$,
$\bar g(z_1,z_2,z_3)=g(x_1,x_2,x_3)$ and ${\bar u}_d(z_1,z_2,z_3)=u_d(x_1,x_2,x_3)$.


Then, using these notations, the rescaled energy takes the form
\begin{eqnarray}
  J'^\e (\hat{u}^\varepsilon,\bar{u}^\varepsilon) &= & \int_{\Omega_+} ( \frac{1}{2} A_+
 ( e(\bar{u}^\varepsilon)) \cdot  e(\bar{u}^\varepsilon) - \bar{f} \cdot \bar{u}^\varepsilon
) \; dz
   - \int_{\bar{\Gamma}_1} \bar{g} \cdot \bar{u}^\varepsilon\;dS  \nonumber \\
   && \ \ \ +  \int_{B} \frac{1}{2} ( \varepsilon^{-1} {\hat K}^{33} (\hat{u}^\varepsilon_{,3}) \cdot
   \hat{u}^\varepsilon_{,3} + 2 {\hat K}^{\alpha 3} (\hat{u}^\varepsilon_{,\alpha})\cdot
   \hat{u}^\varepsilon_{,3}  \label{eq:resc-en}\\
   && \ \ \ \qquad + \varepsilon {\hat K}^{\alpha  \beta} (\hat{u}^\varepsilon_{,\alpha}) \cdot
   \hat{u}^\varepsilon_{,\beta} ) \; dz    \nonumber
\end{eqnarray}
where  a comma is used to denote partial differentiation, $\alpha, \beta \in \{1,2\}$ and
$K^{jl}, j, l = 1, 2, 3,$ are the matrices whose components are
defined by the relations
\begin{equation}\label{eq:ke}
  ({\hat K}^{jl})_{ki} := \hat{A}_{ijkl}.
\end{equation}
In view of the symmetry properties of the elasticity tensor $\hat{A},$ the
matrices ${\hat K}^{jl}$ have the property that ${\hat K}^{jl} = ({\hat K}^{lj})^T, \ j,
l = 1, 2, 3.$
\begin{centering}
\begin{figure}[ht!]
\centerline{\resizebox{9.cm}{!}{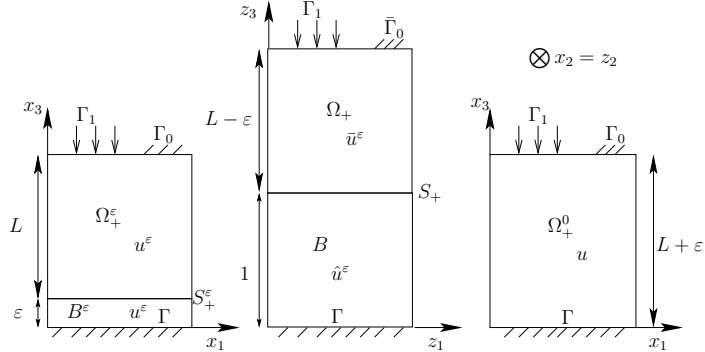}}

  \caption{(a) Initial, (b) rescaled , and  (c) limit configuration of a solid glued to a rigid base. }\label{fig:app}
\end{figure}
\end{centering}
The rescaled equilibrium problem  is formulated as follows:  find the
pair $(\bar{u}^\varepsilon, \hat{u}^\varepsilon ) $
minimizing the  energy (\ref{eq:resc-en}) in the set of displacements
\begin{eqnarray}
    V' = \{ (\bar{u}^\e, \hat{u}^\varepsilon) \in H(\Omega_{+} ;R^3) \times
H(B;R^3) :
\bar{u}^\e = {\bar u}_d\mbox{ on } {\bar \Gamma_1},\qquad\nonumber\\
\ \bar{u}^\e= \hat{u} ^\e\ \ \mathrm{on} \ S_{+}, \hat{u}^\e= 0 \ \ \mathrm{on} \ \Gamma \}.
\end{eqnarray}
where $H(\Omega_{+} ;R^3)$  and $H(B ;R^3)$ are the sets of admissible displacements defined on $\Omega_{+}$ and $B$, respectively.

We assume that the displacements minimizing $ J'^\e$ in $V'$ can be expressed as the sum of
the series
\begin{eqnarray}
\hat{u}^\varepsilon &=& \hat{u}^0 + \varepsilon \hat{u}^1 + \varepsilon^2 \hat{u}^2 +
o(\varepsilon^2) \;, \label{eq:exptil}\\
\bar{u}^\varepsilon &=& \bar{u}^0 + \varepsilon \bar{u}^1 + \varepsilon^2 \bar{u}^2 +
o(\varepsilon^2) \;.\label{eq:expbar}
\end{eqnarray}
Correspondingly, the rescaled energy (\ref{eq:resc-en})
can be written as:
\begin{eqnarray}
J'^\e (\hat{u}^\varepsilon,\bar{u}^\varepsilon)& =&\frac1\e J'^{-1}(\hat{u}^0)+\e^0 J'^0(\hat{u}^0,\bar{u}^0,\bar{u}^1) +\e J'^1(\hat{u}^0,\bar{u}^0,\hat{u}^1,\bar{u}^1,\hat{u}^2)\nonumber \\
 && +\e^2 J'^2(\hat{u}^0,\bar{u}^0,\hat{u}^1,\bar{u}^1,\hat{u}^2,\bar{u}^2,\hat{u}^3) +o(\e^2),
\end{eqnarray}
where
\begin{eqnarray}
  J'^{-1}&:=&J'^{-1}(\hat{u}^0) \\
  &:=& \int_{B} \frac{1}{2}
( {\hat K}^{33} ( \hat{u}^0_{,3}) \cdot \hat{u}^0_{ ,3} ) \; dz, \nonumber \\
 J'^0 &:=& J'^0(\hat{u}^0,\bar{u}^0,\bar{u}^1) \\
&:=& \int_{\Omega_{+}} ( \frac{1}{2} A_{+} (
  e(\bar{u}^0)) \cdot  e(\bar{u}^0) - \bar{f} \cdot \bar{u}^0
) \; dz
   - \int_{\bar{\Gamma}_1} \bar{g} \cdot \bar{u}^0 \;dS + \nonumber \\
   && \ \  + \int_{B} (
{\hat K}^{33} ( \hat{u}^0_{ ,3}) \cdot \hat{u}^1_{,3} + {\hat K}^{\alpha 3} (
\hat{u}^0_{ ,\alpha}) \cdot \hat{u}^0_{,3} ) \; dz,\nonumber  
\end{eqnarray}
\begin{eqnarray}
J'^1&:=&J'^1(\hat{u}^0,\bar{u}^0,\hat{u}^1,\bar{u}^1,\hat{u}^2)\\
&:=& \int_{\Omega_+} ( A_{+} (
  e(\bar{u}^0)) \cdot  e(\bar{u}^1) - \bar{f} \cdot \bar{u}^1
) \; dz
   - \int_{\bar{\Gamma}_1} \bar{g} \cdot \bar{u}^1 \;dS + \nonumber \\
   && \ \  + \int_{B} ( {\hat K}^{33} ( \hat{u}^0_{,3} ) \cdot \hat{u}^2_{,3} + \frac{1}{2}
{\hat K}^{33} ( \hat{u}^1_{,3}) \cdot \hat{u}^1_{,3} ) \; dz + \nonumber \\
&& \ \ + \int_{B} ({\hat K}^{\alpha 3} ( \hat{u}^0_{,\alpha}) \cdot
\hat{u}^1_{,3} + {\hat K}^{\alpha 3} (\hat{u}^1_{ ,\alpha}) \cdot
\hat{u}^0_{ ,3} ) \; dz + \nonumber \\
&&\ \  + \int_{B}  \frac{1}{2} {\hat K}^{\alpha
\beta} ( \hat{u}^0_{ ,\alpha}) \cdot \hat{u}^0_{ ,\beta}  \; dz,\nonumber  \\
J'^2&:=&J'^2(\hat{u}^0,\bar{u}^0,\hat{u}^1,\bar{u}^1,\hat{u}^2,\bar{u}^2,\hat{u}^3)\\
&:=&
\int_{\Omega_+} ( \frac{1}{2} A_{+} (
  e(\bar{u}^1)) \cdot e(\bar{u}^1) - \tilde{f} \cdot \bar{u}^2
) \; dz
   - \int_{\bar{\Gamma}_1} \bar{g} \cdot \bar{u}^2 \;dS + \nonumber \\
   && \ \ +\int_{\Omega_{+}}   A_{+} (
  e(\bar{u}^0)) \cdot e(\bar{u}^2) \; dz + \int_{B}  {\hat K}^{33}(\hat{u}^0_{,3}) \cdot \hat{u}^3_{,3} \; dz + \nonumber \\
  && \ \ + \int_{B}(  {\hat K}^{3 3}( \hat{u}^1_{ , 3}) \cdot
\hat{u}^2_{,3} +  {\hat K}^{\alpha 3}( \hat{u}^0_{,\alpha}) \cdot
\hat{u}^2_{,3} +   {\hat K}^{\alpha 3} (\hat{u}^1_{,\alpha}) \cdot
\hat{u}^1_{,3} ) \; dz + \nonumber \\
&& \ \ + \int_{B}  ( {\hat K}^{\alpha 3} ( \hat{u}^2_{ ,\alpha}) \cdot
\hat{u}^0_{ ,3}\; dz +{\hat  K}^{\alpha \beta} ( \hat{u}^0_{,\alpha}) \cdot
\hat{u}^1_{,\beta} ) \; dz.\nonumber
\end{eqnarray}
As proposed in \cite{LR2010}, we now minimize successively the energies $J'^{-1}$, $J'^0$, $J'^1$, and $J'^2.$

\subsubsection{Minimization of $J'^{-1}$}

The energy $J'^{-1}$ is minimized in the class of displacements
\begin{equation}\label{eq:v'm1}
    V'_{-1} = \{ \hat{u}^0 \in H(B ;R^3) : \hat{u}^0 = 0 \ \ \mathrm{on} \ \Gamma \}.
\end{equation}
Because $\hat{A}$ is a positive definite tensor, the second order tensor
${\hat K}^{33}$ is also positive definite. Therefore, the energy
$J'^{-1}$ is non negative and the minimizers are such that
\begin{equation}
   \hat{u}^0_{, 3} = 0,  \ \ a. e. \ \ \mathrm{in} \
   B,
\end{equation}
which, together with the boundary condition in (\ref{eq:v'm1}), implies
\begin{equation}\label{eq:u0}
   \hat{u}^0 = 0,  \ \ a. e. \ \ \mathrm{on} \
   S_+.
\end{equation}

\subsubsection{Minimization of $J'^{0}$}

Based on (\ref{eq:u0}), the energy
$J'^{0}$ turns out to depend only on $\bar{u}^0$ and it takes the form:
\begin{equation}
J'^0(\hat{u}^0,\bar{u}^0,\bar{u}^1) =
\int_{\Omega_{+}} ( \frac{1}{2} A_{+} (
  e(\bar{u}^0)) \cdot  e(\bar{u}^0) - \bar{f} \cdot \bar{u}^0
) \; dz
   - \int_{\bar{\Gamma}_1} \bar{g} \cdot \bar{u}^0 \;dS.
\end{equation}
In view of (\ref{eq:u0}) and of the continuity of the displacements at the surface $\hat{S}_{+},$ we seek the energy minimizer
in the class of displacements
\begin{equation}
    V'_{0} =\{ \bar{u}^0 \in H(\Omega_{+} ;R^3) : \bar{u}^0 = 0 \ \ \mathrm{on} \ {S}_{+},
  \ {\bar u}^0={\bar u}_d\mbox{ on } {\bar \Gamma}_0\}.\label{eq:v'0}
\end{equation}
Using standard arguments, we obtain the equilibrium equations
\begin{eqnarray}
  \mathrm{div} ( A_+ (e(\bar{u}^0))) + \bar{f} &=& 0 \ \ \ \mathrm{in} \ \Omega_{+}, \label{eq:div0} \\
  A_+ (e(\bar{u}^0))n  &=& {\bar g} \ \ \ \mathrm{on} \ \bar{\Gamma}_1, \label{eq:nat00} \\
A_+ (e(\bar{u}^0))n  &=& 0 \ \ \ \mathrm{on} \ \partial \Omega_{+} \setminus (\bar{\Gamma}_1 \cup {S}_{+}). \label{eq:nat0g}
\end{eqnarray}

\subsubsection{Minimization of $J'^{1}$}

In view of (\ref{eq:u0}), the energy $J'^{1}$ simplifies as follows:
\begin{eqnarray}
J'^1(\hat{u}^0,\bar{u}^0,\hat{u}^1,\bar{u}^1,\hat{u}^2)
&&:= \int_{\Omega_+} ( A_{+} (
  e(\bar{u}^0)) \cdot  e(\bar{u}^1) - \bar{f} \cdot \bar{u}^1
) \; dz \nonumber \\
   &&  - \int_{\bar{\Gamma}_1} \bar{g} \cdot \bar{u}^1 \;dS+ \int_{B} (  \frac{1}{2}
{\hat K}^{33} ( \hat{u}^1_{,3}) \cdot \hat{u}^1_{,3} ) \; dz.
\end{eqnarray}
We minimize this energy
in the class of displacements
\begin{eqnarray}
    V'_{1} &=& \{ (\bar{u}^1,\hat{u}^1) \in H(\Omega_{+} ;R^3) \times  H(B ;R^3)  : \bar{u}^1 = \hat{u}^1 \ \ \mathrm{on} \ {S}_{+}, \nonumber \\
    &&\qquad \hat{u}^1 = 0 \ \ \mathrm{on} \ \Gamma, \ {\bar u}^1=0\mbox{ on } {\bar \Gamma}_0\}. \label{eq:v'1}
\end{eqnarray}
Using (\ref{eq:div0}-\ref{eq:nat0g}), the Euler-Lagrange equations reduce to the following equation:
\begin{equation}\label{eq:min1}
\int_{{S}_{+}} (A_{+} (
  e(\bar{u}^0))n \cdot \bar{\eta}^1) dS + \int_{B} (
{\hat K}^{33} ( \hat{u}^1_{,3}) \cdot \hat{\eta}^1_{,3} ) \; dz = 0,
\end{equation}
where $\bar{\eta}^1, \hat{\eta}^1$ are perturbation of $\bar{u}^1, \hat{u}^1,$ respectively, and they are such that $\bar{\eta}^1= \hat{\eta}^1$
on ${S}_{+}.$ Integrating by parts the second integral
and using the
boundary conditions in (\ref{eq:v'1}), we obtain
\begin{equation}\label{eq:min12}
\int_{{S}_{+}} (A_{+} (
  e(\bar{u}^0))n \cdot \bar{\eta}^1) dS + \int_{{S}_{+}} ({\hat K}^{33} ( \hat{u}^1_{,3}) \cdot \hat{\eta}^1) dS - \int_{B} (
{\hat K}^{33} ( \hat{u}^1_{,33}) \cdot \hat{\eta}^1 ) \; dz  = 0.
\end{equation}
Using the arbitrariness of $\bar{\eta}^1, \hat{\eta}^1,$ we obtain
\begin{eqnarray}
  \hat{u}^1_{,33} &=& 0 \ \ \mathrm{in} \ B, \\
  \hat{u}^1_{,3} &=& - ({\hat K}^{33})^{-1} (A_{+} (
  e(\bar{u}^0))n  \ \ \mathrm{on} \ {S}_{+},
\end{eqnarray}
which, together with the boundary condition on $\Gamma,$ give
\begin{equation}\label{eq:u1}
\hat{u}^1 =  - \Big( ({\hat K}^{33})^{-1} (A_{+} (
  e(\bar{u}^0))n \Big) z_3 \ \ \mathrm{in} \ B.
\end{equation}

\subsubsection{Minimization of $J'^{2}$}

Using the results obtained so far, the energy $J'^{2}$ simplifies as follows:
\begin{equation}\label{eq:j'2}
J'^2(\hat{u}^0,\bar{u}^0,\hat{u}^1,\bar{u}^1,\hat{u}^2,\bar{u}^2,\hat{u}^3) = \int_{\Omega_+} ( \frac{1}{2} A_{+} (
  e(\bar{u}^1)) \cdot e(\bar{u}^1)  dz + \int_{B}    {\hat K}^{\alpha 3} (\hat{u}^1_{,\alpha}) \cdot
\hat{u}^1_{,3}  \; dz
\end{equation}
In view of (\ref{eq:u1})  the second integral is a constant term and it can be dropped in the minimization procedure.
Thus, we minimize the remaining term in the energy
in the class of displacements
\begin{equation}\label{eq:v'2}
    V'_{2} = \{ \bar{u}^1 \in H(\Omega_{+} ;R^3)  : \bar{u}^1 = - \Big( ({\hat K}^{33})^{-1} (A_{+} (
  e(\bar{u}^0))n \Big) \ \ \mathrm{on} \ {S}_{+}\}.
\end{equation}
and we obtain the equilibrium equations:
\begin{eqnarray}
  \mathrm{div} ( A_+ (e(\bar{u}^1))) &=& 0 \ \ \ \mathrm{in} \ \Omega_{+}, \label{eq:div1} \\
A_+ (e(\bar{u}^1))n  &=& 0 \ \ \ \mathrm{on} \ \partial \Omega_{+} \setminus {S}_{+}. \label{eq:nat1g}
\end{eqnarray}

\subsubsection{Limit equilibrium problems}

Due to the continuity of the displacement across the surface $S_+$, we have
$$
\hat u^\e(z_1,z_2,1^-)= \bar u^\e(z_1,z_2,1^+) = u^\e(x_1,x_2,\e).
$$
Note that the same condition is obtained for the stress field along $S_+$.

Using an asymptotic expansion, we have $u^\e(x_1,x_2,\e)=u^\e(x_1,x_2,0^+)+\e u^\e_{,3}(x_1,x_2,0^+)+o(\e)$.

Using relations (\ref{mae}) and the last two equations, we obtain
\begin{eqnarray}
  u^0(x_1,0^+) &=& \bar{u}^0(x_1,1^+), \\
  u^0_{,3}(x_1,0^+) +  u^1(x_1,0^+)&=& \bar{u}^1(x_1,1^+).
\end{eqnarray}

With the help of the above relations, we can rewrite the interface conditions obtained in the asymptotic analysis in terms of the displacement in the deformable adherent. In  summary, we have the following two equilibrium problems:
\begin{eqnarray}
&& (P_0) \ \ \left \{ \begin{array}{ll}
           \mathrm{div} ( A_+ (e(u^0))) + \bar{f} = 0 & \mathrm{in} \ \Omega^0_{+}, \\
            A_+ (e(u^0))n  = g  & \mathrm{on} \ \Gamma_1, \\
            A_+ (e(u^0))n  = 0 & \mathrm{on} \ \partial \Omega^0_{+} \setminus (\Gamma_1 \cup \Gamma) \\
            u^0 = 0 & \mathrm{on} \ \ \Gamma, \
          \end{array} \right. \label{plane0}\\
&& (P_1) \ \ \left \{ \begin{array}{ll}
           \mathrm{div} ( A_+ (e(u^1)) ) = 0 & \mathrm{in} \ \Omega^0_{+}, \\
            A_+ (e(u^1))n  = 0 & \mathrm{on} \ \partial \Omega^0_{+} \setminus \Gamma \\
            u^1 = - \Big( ({\hat K}^{33})^{-1} (A_{+} (
  e({u}^0))n \Big) - u^0_{,3} & \mathrm{on} \ \ \Gamma. \
          \end{array} \right.  \label{plane1}
\end{eqnarray}

In the remaining of the paper, the problem ($P_0$) will be referred to as the ‘‘problem at the order zero", because its unknown is $u^0$.
Similarly,  the problem ($P_1$) will be referred to as the ‘‘problem at the first order", because its unknown is $u^1$ and the displacement
 $u^0$ is considered as given and calculated by using  ($P_0$).

\subsection{Plane strain problem}
We consider the case of plane strain in the plane $(x_1,x_2)$, where the interface between the glue and the adhesive
is a line orthogonal to the direction $e_2$.
After adopting natural notations, it is obvious to obtain, by  implementing the same kind of technique used in the previous section, the two following problems:

\begin{eqnarray}
&& (P_0) \ \ \left \{ \begin{array}{ll}
           \mathrm{div} ( A_+ (e(u^0))) + {f} = 0 & \mathrm{in} \ \Omega^0_{+}, \\
            A_+ (e(u^0))n  = g  & \mathrm{on} \ \Gamma_1, \\
            A_+ (e(u^0))n  = 0 & \mathrm{on} \ \partial \Omega^0_{+} \setminus (\Gamma_1 \cup \Gamma) \\
            u^0 = 0 & \mathrm{on} \ \ \Gamma, \
          \end{array} \right. \\
&& (P_1) \ \ \left \{ \begin{array}{ll}
           \mathrm{div} ( A_+ (e(u^1)) ) = 0 & \mathrm{in} \ \Omega^0_{+}, \\
            A_+ (e(u^1))n  = 0 & \mathrm{on} \ \partial \Omega^0_{+} \setminus \Gamma \\
            u^1 = - \Big( ({\hat K}^{22})^{-1} (A_{+} (
  e({u}^0))n \Big) - u^0_{,2} & \mathrm{on} \ \ \Gamma. \
          \end{array} \right.
\end{eqnarray}


\subsection{Case of the gluing of two elastic bodies}

Let us now consider the adhesive bonding of two linear elastic bodies
 satisfying the plane strain hypothesis.

We extend the notation used before, and we define the following domains

 \begin{itemize}
\item $B^\e=\{ (x_1,x_2)\in \Omega:\ |x_2|<\frac\e2  \}$ (the glue);

\item $\Omega^\e_{\pm} =\{ (x_1,x_2)\in \Omega:\  \pm x_2>\frac\e2  \}$;

\item $S^\e_\pm=\{ (x_1,x_2)\in \Omega:\ x_2=\pm \frac{\e}{2}  \}$;

\item $\Gamma=\{ (x_1,x_2)\in \Omega:\ x_2=0  \}$ (the interface);

\item $\Omega_\pm =\{ (x_1,x_2)\in \Omega:\  \pm x_2>\frac12  \}$;

\item $B=\{ (x_1,x_2)\in \Omega:\  |x_2|<\frac12  \}$;

\item $S_\pm=\{ (x_1,x_2)\in \Omega:\ x_{\pm} =\pm \frac12  \}$;

\item $\Omega^0_\pm =\{ (x_1,x_2)\in \Omega:\  \pm x_2>0  \}$.
\end{itemize}



\begin{figure}[!ht]
\centerline{\resizebox{9.cm}{!}{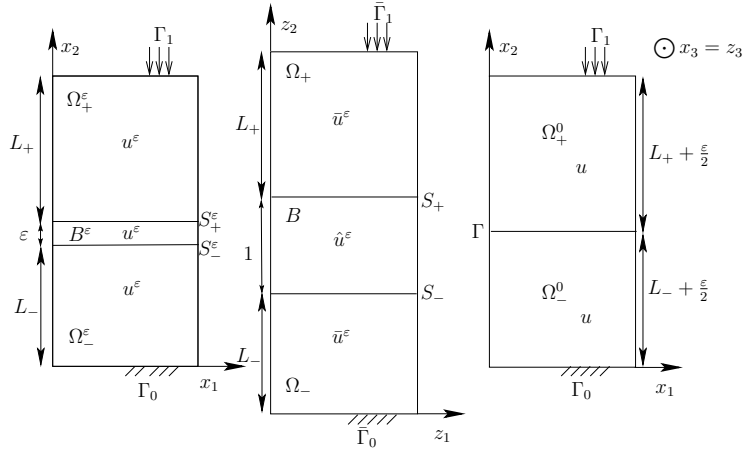}}
\caption{Geometry of the interphase/interface problem. Left: the initial
 problem with an interphase of thickness $\e$ -- Middle: the rescaled problem with interphase height equal to 1 -- Right: the
limit interface problem.}\label{fig1a}
\end{figure}

The methodology and the notations used here are similar to the ones used in the
 previous section.
The main differences are:
\begin{itemize}
\item The introduction of a jump of the stress vector at order 0 and 1;
\item The displacement along the interface is replaced by a jump of the displacement across
the interface between the two bodies;
\item The minimization of $J'^1$ leads to concentrated forces at the edges of the interface.
\end{itemize}

More precisely, the problem at order 0 becomes

\begin{equation}\label{eqeq0}
\left\{\begin{array}{ll}
\mbox{div} \sigma^0 +{ f}=0 \qquad &\mbox{in }\Omega^0_\pm \\
\sigma^0 n=g & \mbox{on } \Gamma_1 \\
u^0=u_d& \mbox{on } \Gamma_0 \\
\sigma^0=A_\pm e(u^0) &\mbox{in }\Omega^0_\pm \\
\lbrack u^0\rbrack=0& \mbox{on } \Gamma\\
\lbrack\sigma^0 n\rbrack=0& \mbox{on } \Gamma
\end{array}\right.
\end{equation}

where $\lbrack f \rbrack (x_1)= f(x_1,0^+)-f(x_1,0^-)$.
The problem at order 1 becomes

\begin{equation}\label{eqeq1}
\left\{\begin{array}{ll}
\mbox{div} \sigma^1 =0 \qquad &\mbox{in }\Omega^0_\pm \\
\sigma^1 n=0 & \mbox{on } \Gamma_1 \\
u^1=0& \mbox{on } \Gamma_0 \\
\sigma^1=A_\pm e(u^1) &\mbox{in }\Omega^0_\pm \\
\lbrack u^1\rbrack=C_1(\sigma^0 n)+C_2(u^{0})_{,1}-\frac12 (u^0(x_1, 0^+)+u^0(x_1,0^-))& \mbox{on } \Gamma\\
\lbrack \sigma^1 n\rbrack=C_3 (\sigma^0 n)_{,1}+C_4(u^0)_{,11} & \mbox{on } \Gamma\\
\sigma^1 e_1=F & \mbox{on } \partial \Gamma\\
\end{array}\right.
\end{equation}

where
$$
C_1=\left(\begin{array}{cc}
\frac1{\hat \mu} & 0 \\
0 & \frac1{\hat \lambda + 2 \hat \mu}
\end{array}\right)
\qquad
C_2=\left(\begin{array}{cc}
0 & -1 \\
-\frac{\hat \lambda}{\hat \lambda+ 2\hat \mu} & 0
\end{array}\right),
$$

$$
C_3=
\left(\begin{array}{cc}
0 & -\frac{\hat \lambda}{\hat \lambda+ 2\hat \mu}  \\
-1  & 0
\end{array}\right)
\qquad
C_4=
\left(\begin{array}{cc}
-4\hat \mu \frac{\hat \lambda+\hat \mu}{\hat \lambda+ 2\hat \mu} & 0 \\
0 & 0
\end{array}\right),
$$

and the localized forces which appear on the lateral boundary of the thin layer are given by
\begin{equation}
F=C_3 (\sigma^0 e_1)+C_4 (u^0)_{,1}
\end{equation}
\end{itemize}

Let us notice that this term appears naturally in this method, and has been first observed
by one another technique in \cite{Z2000} (see also \cite{LR2007, LZ2010}).

%
%
%

\section{Numerical method}

In this paragraph, we present the numerical method developed to solved problem
(\ref{eqeq1}). 
The generic problem associated to this problem can be written
\begin{equation}\label{eqgen}
\left\{\begin{array}{ll}
\mbox{div} \sigma(u) =0 \qquad &\mbox{in }\Omega^0_\pm \\
\sigma(u) n=0 & \mbox{on } \Gamma_1 \\
u=0& \mbox{on } \Gamma_0 \\
\sigma=A_\pm e(u) &\mbox{in }\Omega^0_\pm \\
\lbrack u\rbrack=D & \mbox{on } \Gamma\\
\lbrack\sigma(u) n\rbrack=G& \mbox{on } \Gamma\\
\end{array}\right.
\end{equation}

where $D$ and $G$ are given functions, provided by the solutions $u^0$ and $\sigma^0$ of problem (\ref{eqeq0}) at order 0.
Note the solution of problem at order 0 (\ref{eqeq0}) is very classic and can be solved using a classical finite element method.
In the following, we will denote the restriction of $u$ on $\Omega^0_+$ (resp. $\Omega^0_-$) by $u^+$ (resp. $u^-$).

First, we write the variational formulation of the four first equations of (\ref{eqgen}) both in $\Omega^0_-$ and $\Omega^0_+$,
that leads, after an integration by parts, to
\begin{equation}\label{vf1}
\int_{\Omega^0_\pm}  A_\pm e(u^\pm) \cdot e(v^\pm) dx-\int_{\partial \Omega^0_\pm} (A_\pm e(u^\pm))n^\pm\cdot v^\pm dS
= 0
\end{equation}
for $v^\pm\in \{v\in H^1(\Omega):v=0\mbox{ on } \partial \Gamma_0 \}$. 
Then, introducing the boundary conditions, we obtain

$$
\int_{\Omega^0_\pm}  A_\pm e(u^\pm)\cdot e(v^\pm) dx-\int_{\Gamma} (A_+ e(u^+))n^+\cdot v^+ dS
-\int_{\Gamma} (A_- e(u^-))n^-\cdot v^- dS
=0.
$$

We now choose the normal $n$ equal to the outward normal of $\Omega^0_-$ ($n=n^-=-n^+$), and we denote
$$
I=\int_{\Gamma} (A_+ e(u^+))n\cdot v^+ dS
-\int_{\Gamma} (A_- e(u^-))n\cdot v^- dS.
$$
Then, using the jump condition $ (A_+ e(u^+))n=(A_- e(u^-))n +G$,
we have
\begin{equation}\label{I1}
I=\int_{\Gamma} (A_- e(u^-))n\cdot (v^+ - v^-) dS + \int_\Gamma G v^+\ dS.
\end{equation}

Similarly, writing the jump condition as $ (A_- e(u^-))n=(A_+ e(u^+))n -G$,
we also have

\begin{equation}\label{I2}
I=\int_{\Gamma} (A_+e(u^+))n\cdot (v^+ - v^-) dS + \int_\Gamma G v^-\ dS.
\end{equation}
In order to have a symmetric variational formulation, we consider the half sum of
(\ref{I1}) and (\ref{I2}):
$$
I=\int_\Gamma \frac12 \left[ (A_+e(u^+))n + (A_- e(u^-))n  \right]\cdot (v^+ - v^-) dS +
\int_{\Gamma} G \left(\frac{v^+ + v^-}{2} \right)\ dS.
$$

Again, in order to have a fully symmetric formulation, we need to add in the left hand side of
equation (\ref{vf1}) the term
$$
\int_{\Gamma}(u^+-u^-) \cdot \frac12 \left[ (A_+e(v^+))n + (A_- e(v^-))n  \right]dS
$$
and we use the fact that $u^+-u^-=D$ on $\Gamma$.

Finally, we have the weak formulation
\begin{eqnarray}
\int_{\Omega^0_+\cup \Omega^0_-}  A_\pm e(u^\pm) \cdot e(v^\pm) dx &+&
\int_\Gamma \left(<A e(u) n>\cdot [v]+[u]\cdot <Ae(v)n>\right) dS = \nonumber\\
&-&\int_{\Gamma}G\cdot <v> dS + \int_{\Gamma} D \cdot <A e(v)n> dS,\label{vf2}
\end{eqnarray}
for all $v\in \{ H^1(\Omega):\ \gamma(v)=0\mbox{ on } \partial \Omega\backslash \Gamma\}$, where
$<\cdot>$ denotes the average of the value of the function on the both sides of the interface $\Gamma$: $<f>=\frac12( f^++f^-)$.
\medskip

This formulation, known as the Nitsche's method \cite{N74} is not stable. It is then necessary
to add a stabilization term such as $\displaystyle \frac{\beta}{h}\int_\Gamma [u]\cdot[v] dS$, where $h$ is the size of the smallest
element of the finite element discretization of $\Omega^0_\pm$ considered, and $\beta>0$ is a fixed number that must be sufficiently large to ensure the
stability of the method (see \cite{BHS10, DGHV06, S95} for the complete study of this method and for a priori and
a posteriori error estimates in the case $D= 0$).

Let us notice that this method is formally equivalent to the use of Lagrange multipliers to enforce the jump
conditions (see \cite{BBM92,BH92, BHS10}), but it takes its advantage on the fact that the Nitsche's method does not increase the
number of unknowns.
\medskip

Unfortunately, this method does not work properly to solve the problem (\ref{eqgen}) as soon as $D\neq 0$. To overcome
this difficulty, we split the problem (\ref{eqgen}) into two parts. More precisely, we write $u^\pm=w^\pm+z^\pm$ where
the unknowns $z^+$ and $z^-$ solve the problems

\begin{equation}\label{eqgen2}
\left\{\begin{array}{ll}
\mbox{div } \sigma(z^\pm) =0 \qquad &\mbox{in }\Omega^0_\pm \\
\sigma(z^\pm) n=0 & \mbox{on } \Gamma_1 \\
z^\pm=0& \mbox{on } \Gamma_0 \\
\sigma(z^\pm)=A^\pm e(z^\pm) &\mbox{in }\Omega^0_\pm \\
z^\pm=\pm \frac12 D & \mbox{on } \Gamma\\
\end{array}\right.
\end{equation}
and then, since $[w]=w^+-w^-=[u]- z^+  z^-=(1-\frac12 -\frac12) D=0$,
$w^\pm$ solve

\begin{equation}\label{eqgen3}
\left\{\begin{array}{ll}
\mbox{div } \sigma(w^\pm) =0 \qquad &\mbox{in }\Omega^0_\pm \\
\sigma(w^\pm) n=0 & \mbox{on } \Gamma_1 \\
w^\pm=0& \mbox{on } \Gamma_0 \\
\sigma(w^\pm)=A^\pm e(w^\pm) &\mbox{in }\Omega^0_\pm \\
\lbrack w\rbrack=0 & \mbox{on } \Gamma\\
\lbrack\sigma(w)n\rbrack=G- \lbrack \sigma(z)n\rbrack & \mbox{on } \Gamma\\
\end{array}\right.
\end{equation}

The two first problems defined in (\ref{eqgen2}) both in $\Omega^0_+$ and $\Omega^0_-$ are standard and can be solved simultaneously using a standard
finite element method. The problem (\ref{eqgen3}) is solved using the Nitsche's method developed above.

\section{Numerical results for an elastic body glued to a rigid base}

In this paragraph, we consider a 2D solid composed of  an aluminum adherent and an epoxy resin interphase,
that glues the structure to a rigid base.

The mechanical coefficients of the materials are the following :

\begin{itemize}
\item In the adhesive (Epoxy resin): $\hat E= 4 GPa$, $\hat \nu=0.33$.

\item In the adherent (Aluminium): ${E}= 70 GPa$, $\nu=0.33$.
\end{itemize}

This application was intentionally selected in order to introduce a significative  difference between the elastic moduli of the adherent and those of the adhesive materials.
\medskip

\begin{figure}[!ht]
\centerline{\resizebox{6.cm}{!}{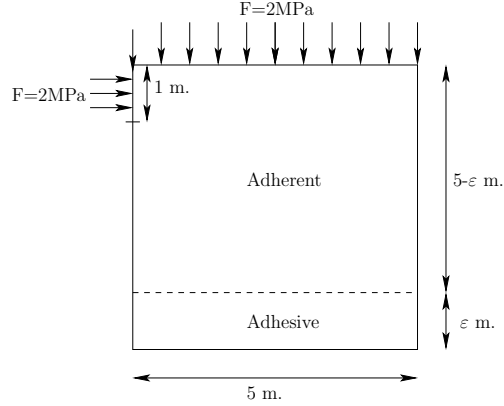}}
\caption{Geometry of the problem ($\e=0$ for the interface problem)}
\label{fig01}       
\end{figure}

The geometry of the problem is provided in figure \ref{fig01}.
The meshes are realized using the GMSH software developed by \cite{gmsh}.
The finite element computations are made with the MATLAB\textsuperscript{\textregistered} software.

The computations are first realized for the interphase problem with various
values of the thickness of the interphase. The values of the jumps of the displacement
and of the stress components across the interphase are then computed.
Independently, the interface problem at order 0 (equations
(\ref{plane0})) is solved numerically. Then, the jumps across the interface at order 1 are
computed using the corresponding equations in (\ref{plane1}), and they are compared with
the jumps across the interphase.
\bigskip

We can notice that, to compute  the jump  in the displacement using (\ref{plane1}), one needs to numerically compute the
derivative of the displacement 
on the boundary. In order to make the computation in a suitable way,
it is necessary to use at least quadratic finite elements. In the numerical experiments below,
we use the quadratic T6 finite elements.

\subsection{Jump $[u]$ across the interface}

\begin{figure}[!ht]
\centering
\includegraphics[height=5.cm]{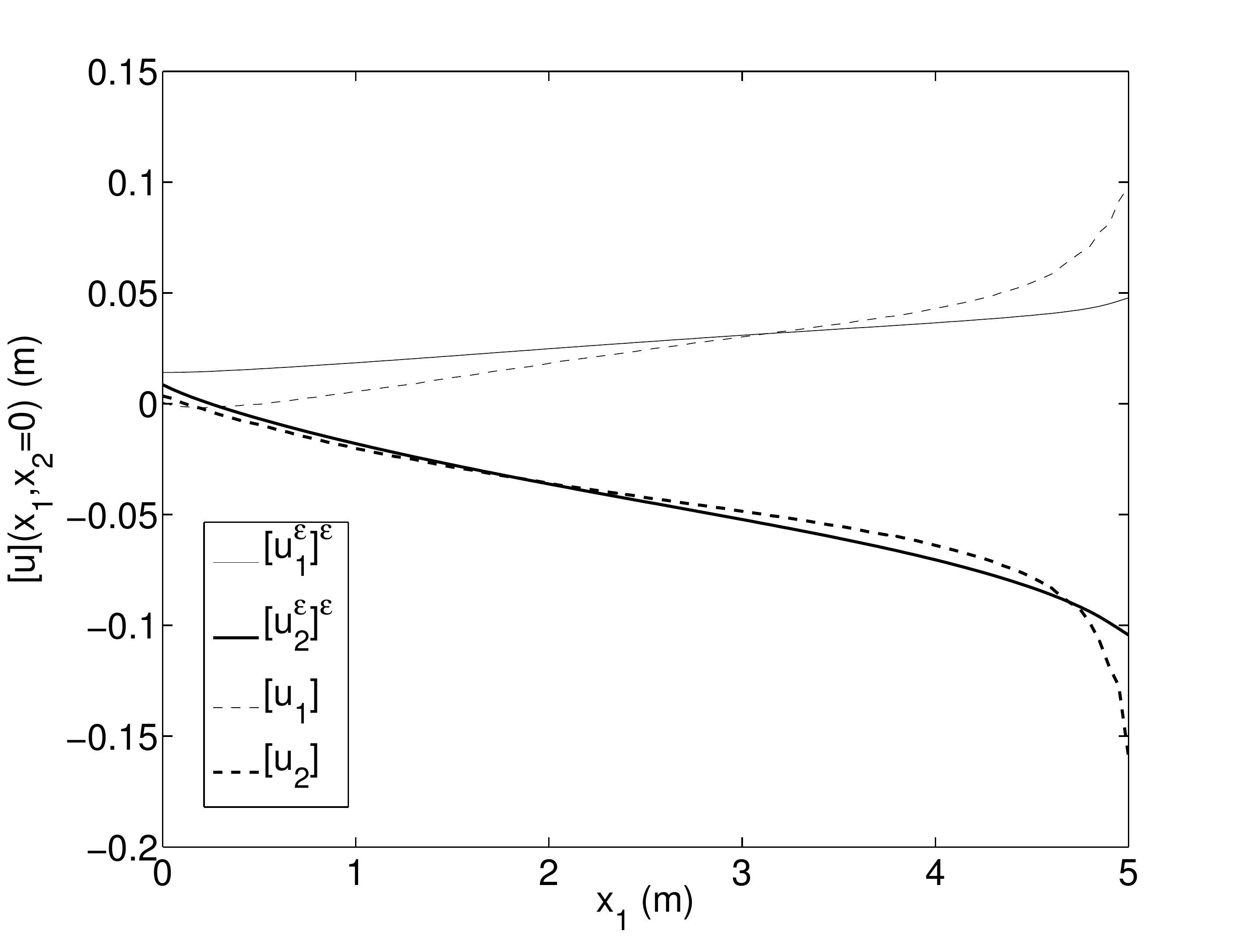}
\includegraphics[height=5.cm]{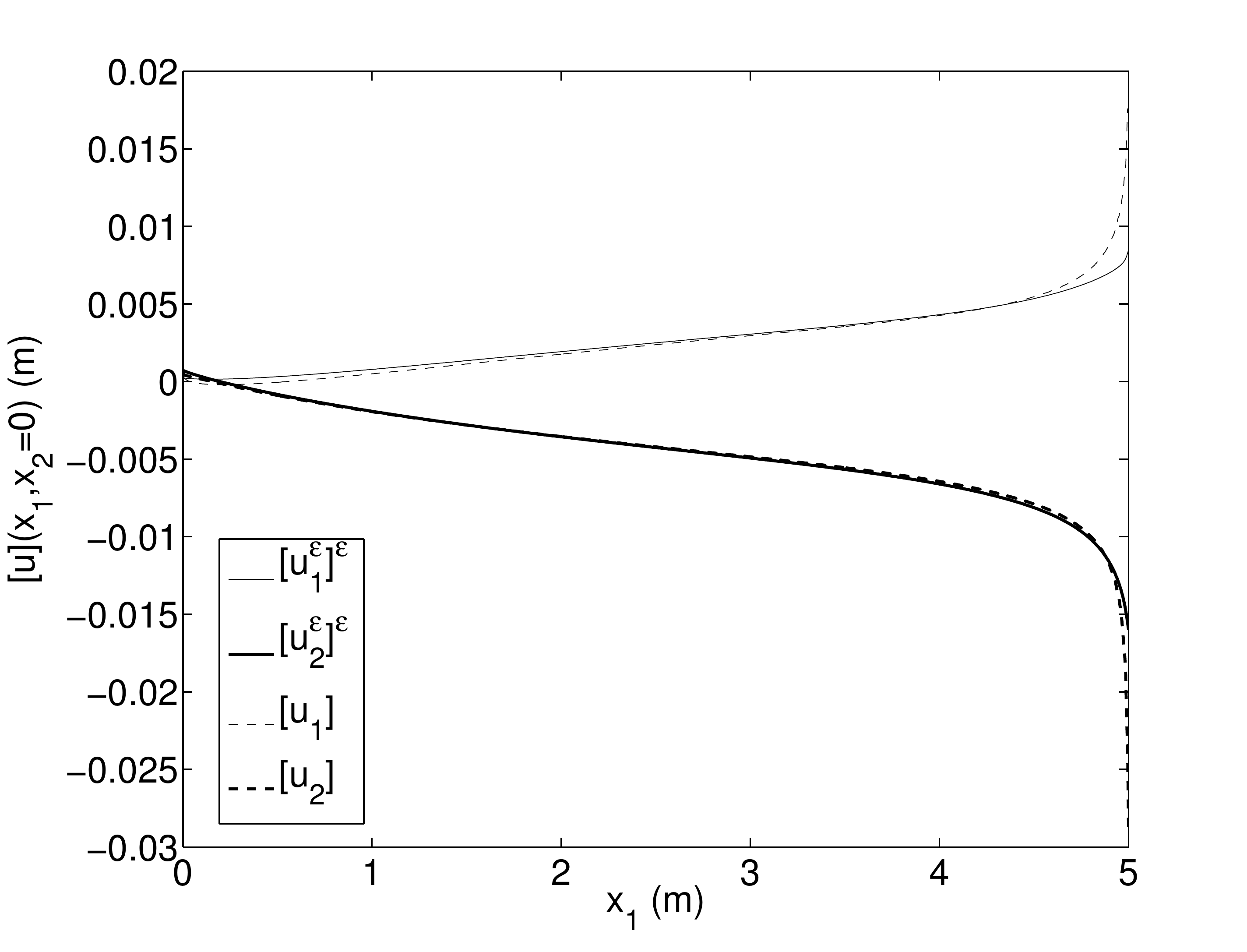}\\
\includegraphics[height=5.cm]{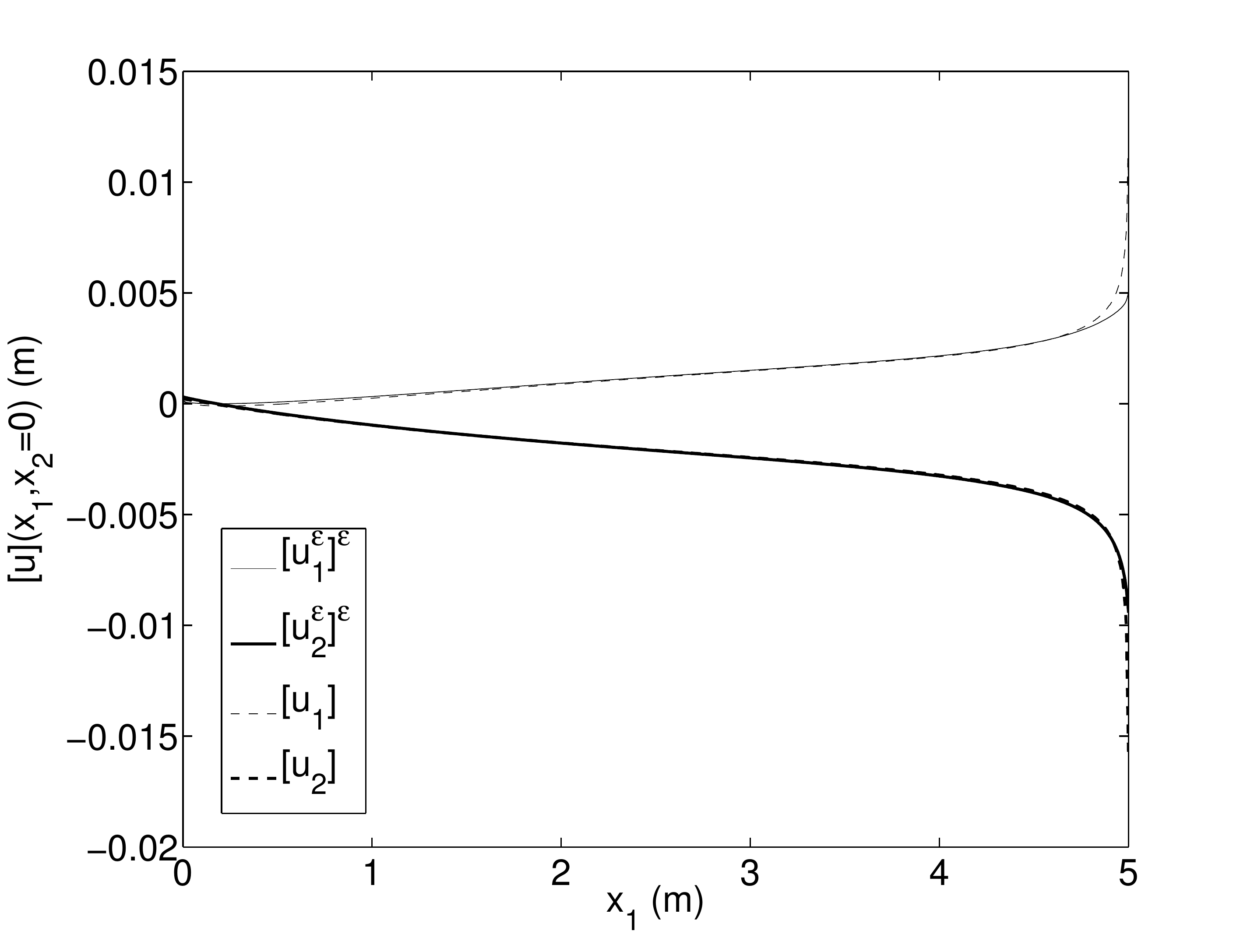}
\includegraphics[height=5.cm]{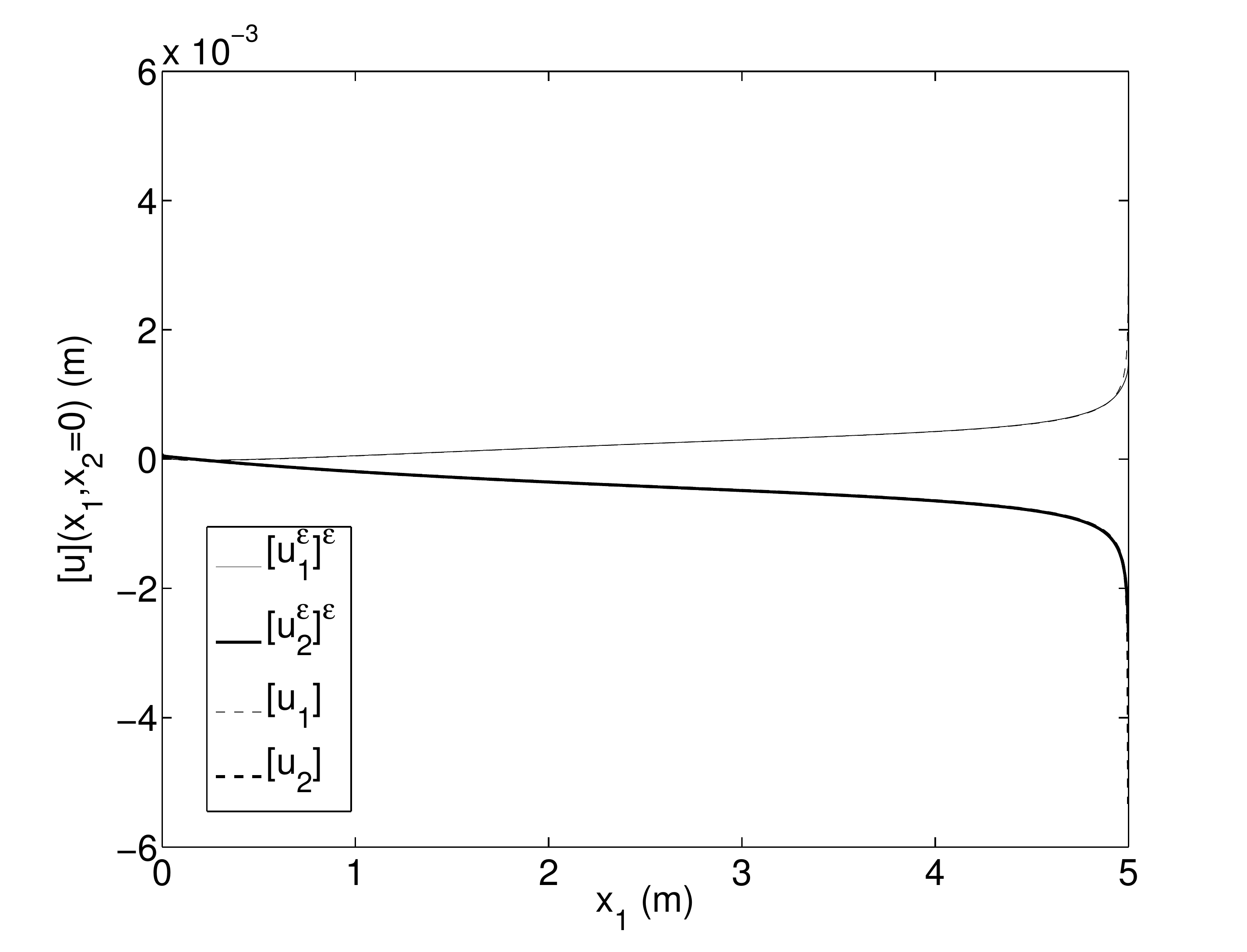}
\caption{Exemple 1 -- Jump in the displacement $[u](x_1,x_2=0)$ (m) along the interface: $\e=0.1$ m (left top), $\e=0.01$ m (right top), $\e=0.005$ m (left bottom)
and $\varepsilon=0.001$ m (right bottom)}
\label{fig29}       
\end{figure}

In figure \ref{fig29}, we present comparative plots of displacement amplitudes $u_1$ and
$u_2$ across
the interface for various values of the thickness $\e$.
More precisely, since the displacement of the rigid base is vanishing, we compare the displacement
$u^\e(x_1=\e,x_2)$, denoted $\lbrack u^\e_i\rbrack$, $i=1,2$ on figure (\ref{fig29}) and computed using the real
geometry of the adhesive, with the displacement $u(x_1=0^+,x_2)=u^0(x_1=0^+,x_2)+\e u^1(x_1=0^+,x_2)$, denoted
$\lbrack u_i\rbrack$, $i=1,2$ and computed
using equations   (\ref{plane0}) and (\ref{plane1}). 
t

We can observe that, as expected, when $\e=0.1$ m the difference between the jumps across the interphase and the jumps
across the interface are relatively large (with a maximum relative error of about 30\%). For smaller values of the thickness $\e$ of the interphase, the difference between
the results obtained using the interphase problem and those obtained using the interface approximation is negligible. More precisely,
the maximum  relative error is close to 1\% for $\e=0.01$ m, $0.6\%$ for $\e=0.005$ m and $0.2\%$ for $\e=0.001$ m.

\subsection{Jump $[\sigma_{12}]$ across the interface}

\begin{figure}[!ht]
\centering
\includegraphics[height=4.6cm]{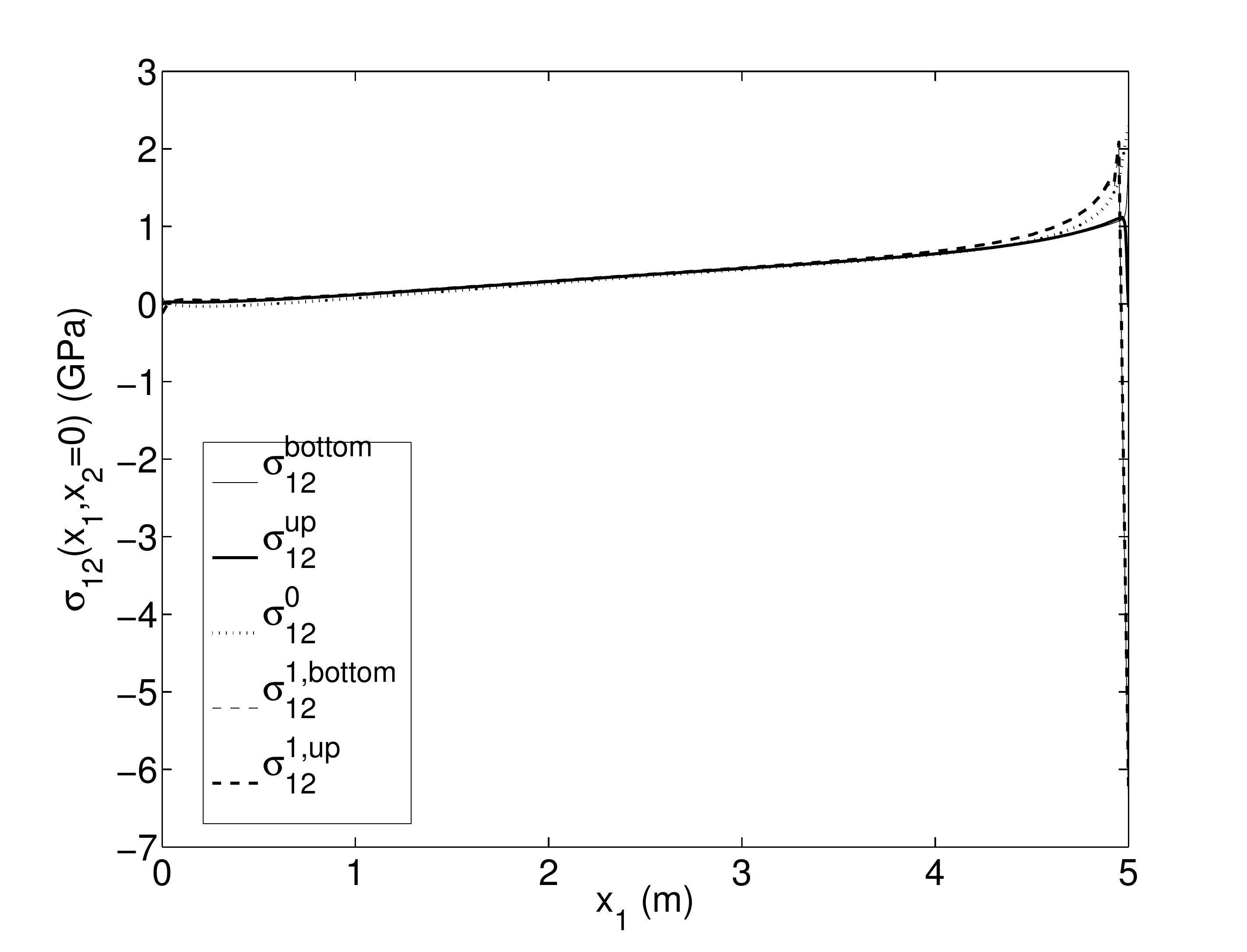}
\includegraphics[height=4.6cm]{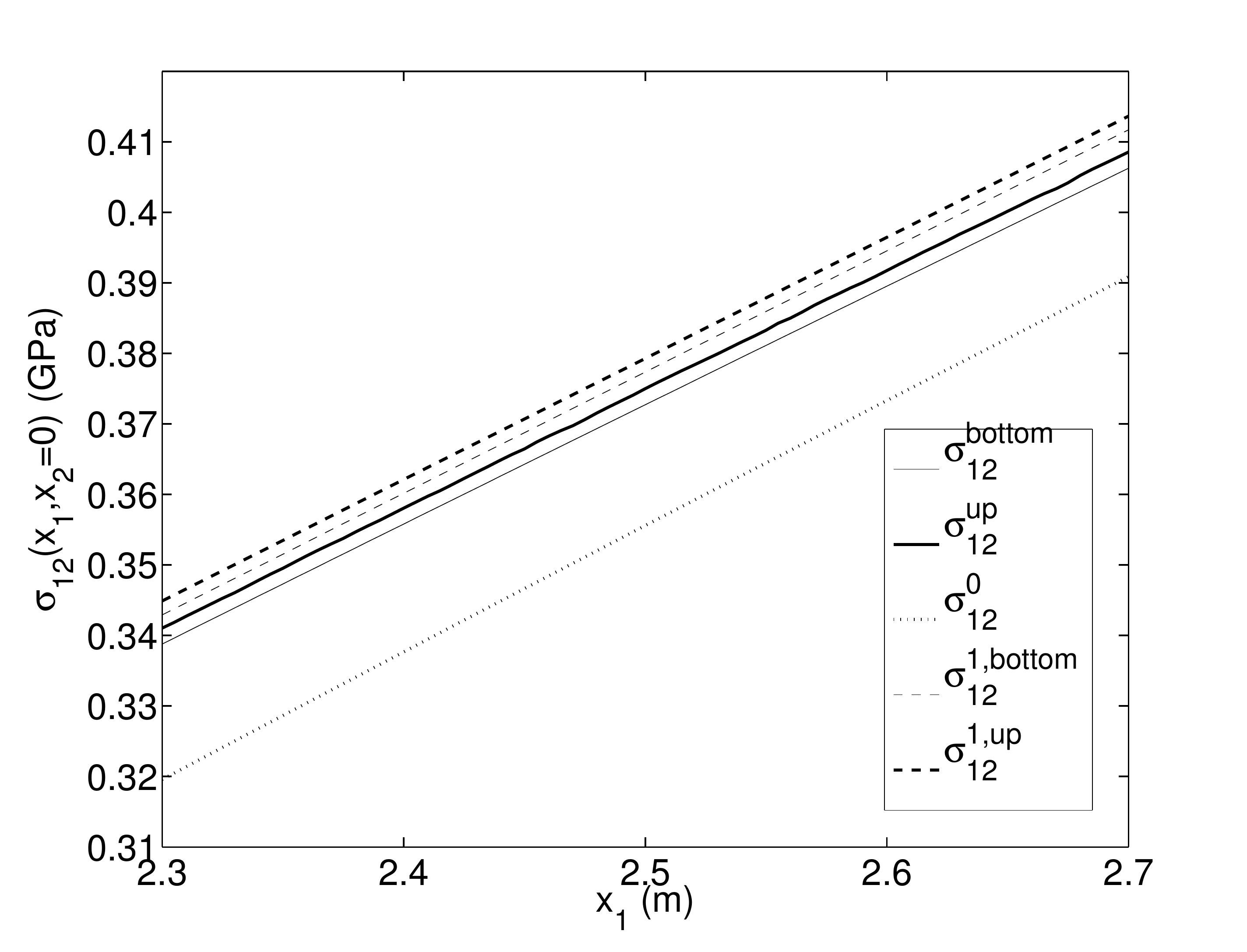}
\caption{Exemple 1 ($\varepsilon=0.01$ m) -- Stress $\sigma_{12}(x_1,x_2=0)$ (GPa) at the bottom of the elastic adherent (up) and on the rigid base (bottom) computed with the various approximations  (zoom on the right)}
\label{fig24b}       
\end{figure}

\begin{figure}[!ht]
\centering
\includegraphics[height=4.6cm]{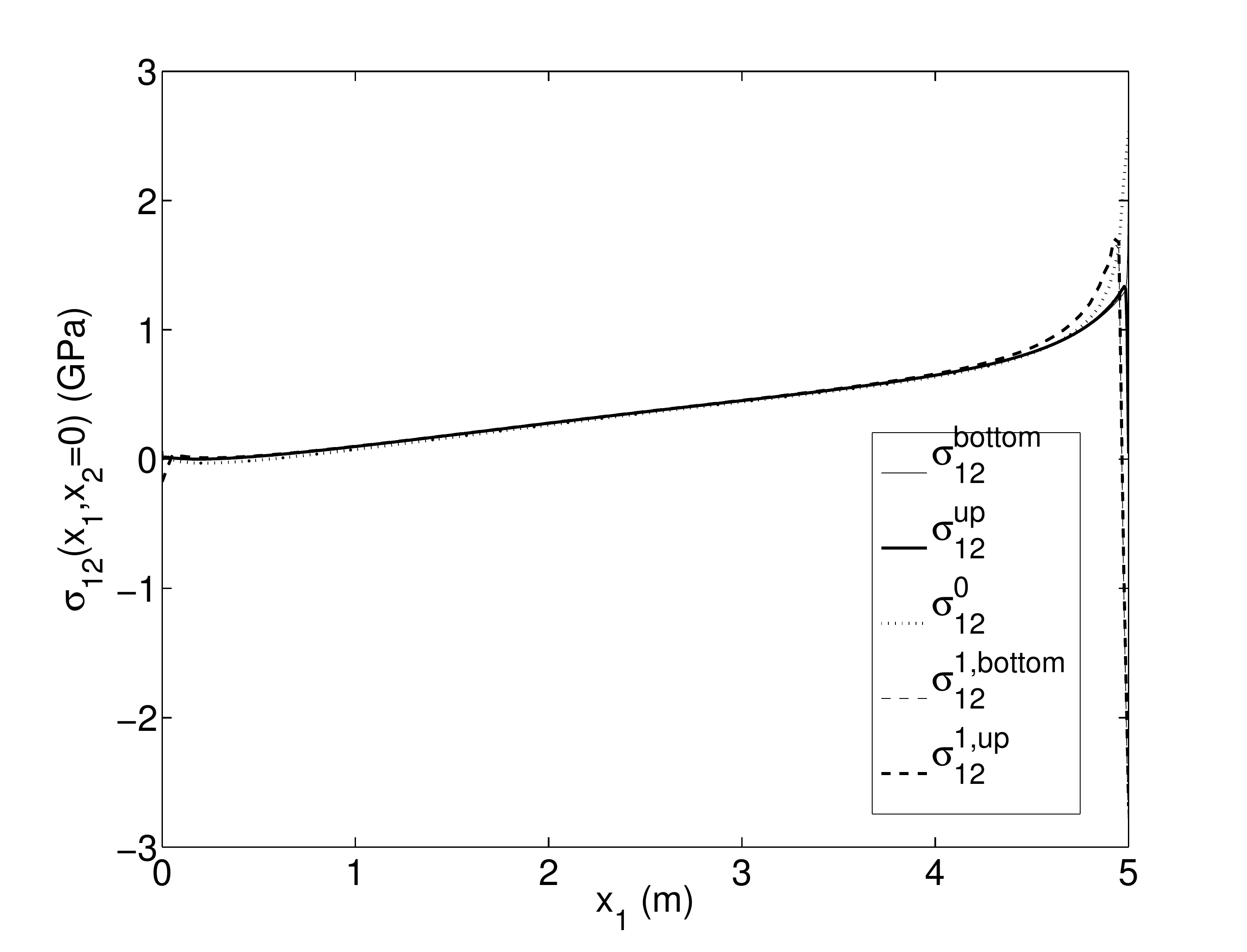}
\includegraphics[height=4.6cm]{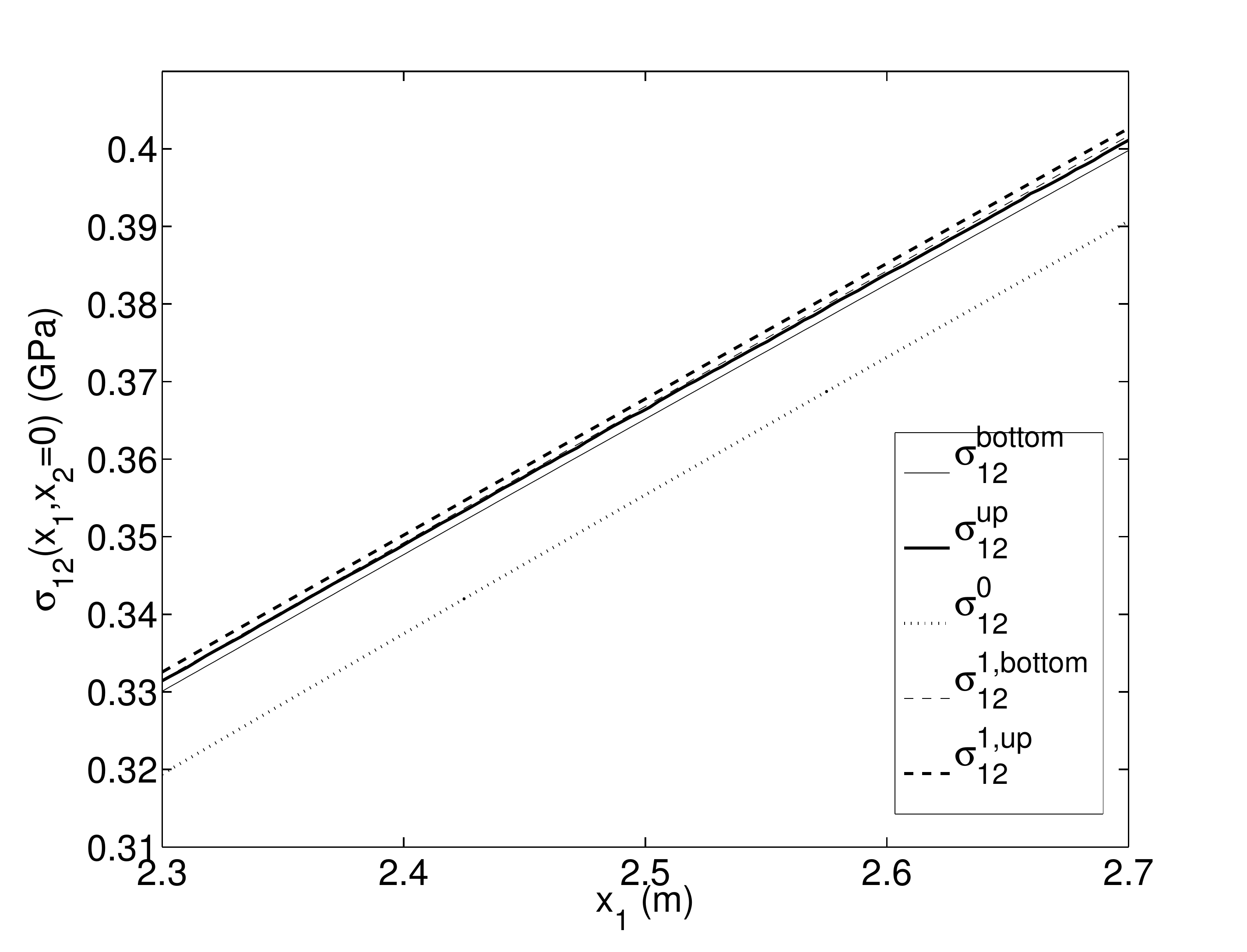}
\caption{Exemple 1 ($\varepsilon=0.005$ m) -- Stress $\sigma_{12}(x_1,x_2=0)$ (GPa) at the bottom of the elastic adherent (up) and on the rigid base (bottom) computed with the various approximations  (zoom on the right)}
\label{fig27b}       
\end{figure}

\begin{figure}[!ht]
\centering
\includegraphics[height=4.6cm]{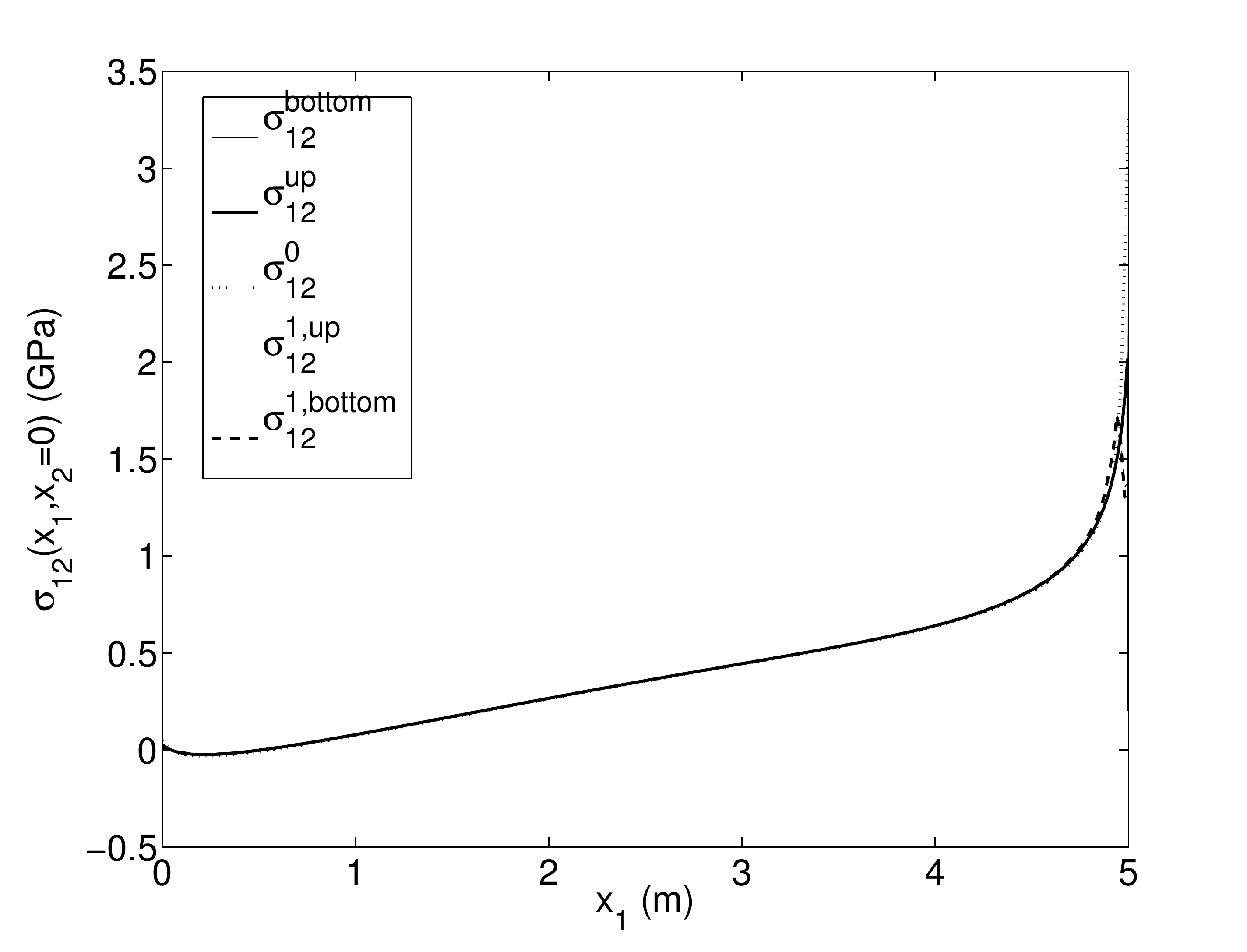}
\includegraphics[height=4.6cm]{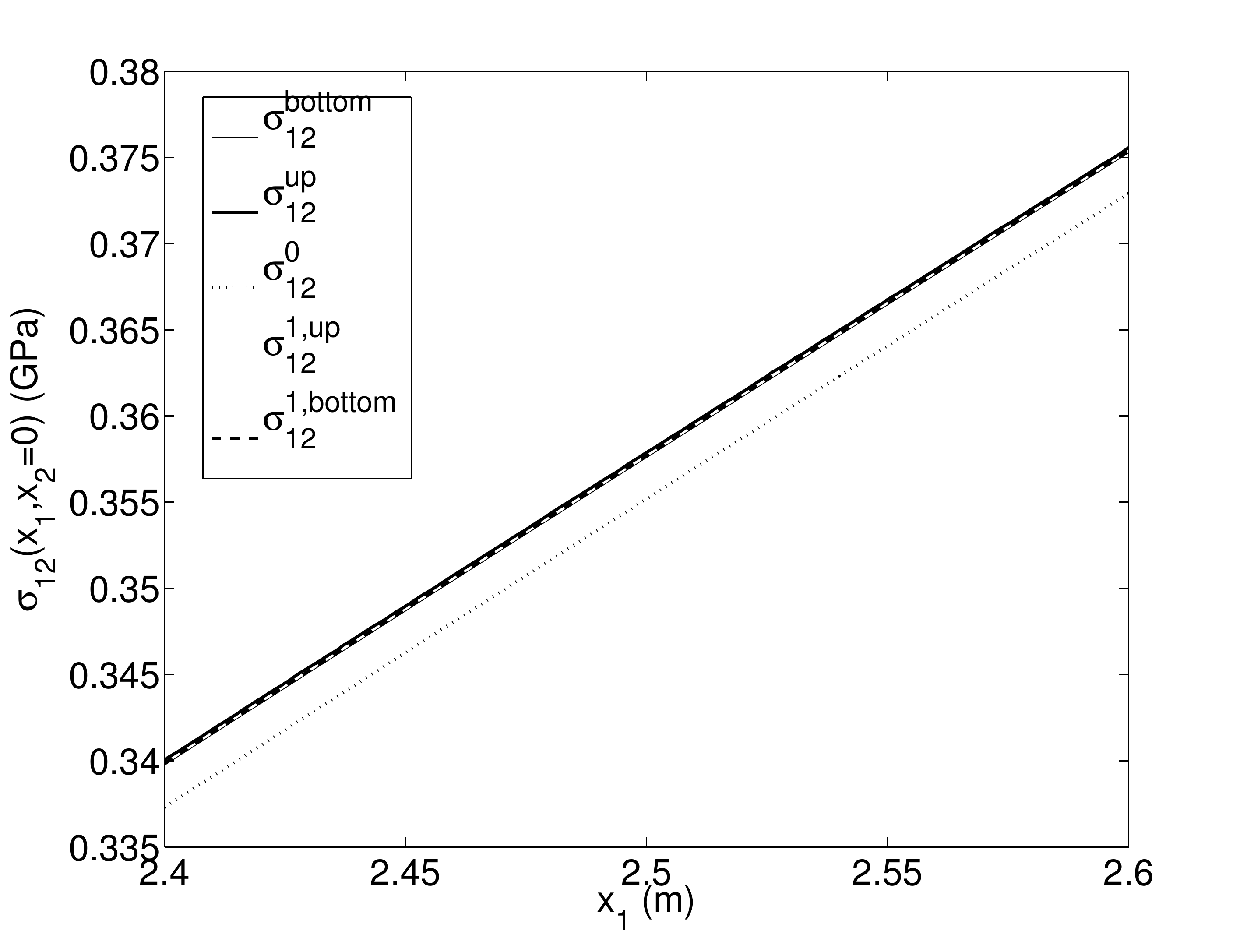}
\caption{Exemple 1 ($\varepsilon=0.001$ m) -- Stress $\sigma_{12}(x_1,x_2=0)$ (GPa) at the bottom of the elastic adherent (up) and on the rigid base (bottom) computed with the various approximations  (zoom on the right)}
\label{fig30b}       
\end{figure}

In this paragraph, we present a comparison between the traction amplitude $\sigma_{12}$  at the top of the interphase/interface and
at the bottom of the interphase/interface computed  for  the original interphase problem and for the approximated (at order 1)  problem with the interface.
They are respectively referred as $\sigma^{up}$ and $\sigma^{bottom}$ on figures (\ref{fig24b}) to (\ref{fig30b}). In the comparison, the thickness $\e$ ranges from 0.01 m to 0.001 m.
We also present the traction amplitude computed  at order 0 for the approximated  problem. Since the traction is continuous
across the interface, i.e. $[\sigma_{12}]=0$, the traction amplitude takes the same value at the top and at the bottom of the interface.

The case $\e=0.1$ m is not presented here because the difference between the case with the interphase
 and the case with the interface is large.

We can observe that the stress $\sigma_{12}$ computed using the interphase problem numerically converges to
the stress $\sigma_{12}$ computed using the interface problem when
$\e$ tends to 0.

One can also observe that the traction amplitude calculated at order 0 converges much slower than the traction amplitude calculated at order 1.
\medskip

In conclusion, it seems that we can replace the interphase behavior by the interface law at order 1, for a thickness of glue lower
than 0.01 m, i.e. less than 1\% of the dimension of the structure.

\subsection{Time computing}

In this paragraph, we present  the time computing necessary to obtain the solutions
of the problems considered in the previous section.

\begin{table}[ht!]
\begin{center}
\begin{tabular}{lllll}
\hline
Thickness & Number of & Number of &  Number of & CPU time \\
$\e$ (m) & nodes & elements & degrees of freedom & (sec.)\\
\hline
$0.1$ & 2,582 & 4,966  & 4,962 & 6.4\\
$0.01$ & 28,172 & 55,166 &  54,342 &  48\\
$0.005$ & 39,486 &  77,432 & 76,304 & 66\\
$0.001$ & 150,054 &  294,831 & 290,106 & 274\\
\hline
\end{tabular}
\caption{Mesh properties (T3 finite elements) and times computing for the interphase problem and various values of $\e$}
\end{center}
\end{table}

\begin{table}[ht!]
\begin{center}
\begin{tabular}{ccccc}
\hline
Thickness & Number of & Number of &  Number of & CPU time \\
$\e$ (m) & nodes & elements & degrees of freedom & (sec.)\\
\hline
$0.1$ & 731 & 1,363 & 1,410 & 5.2 \\
$0.01$ & 10,887 & 5338 & 21,372 & 22\\
$0.005$ & 12,511 & 6,138 &  24,620 & 24\\
$0.001$ & 13,417 & 6,586 & 26,332 & 25 \\
\hline
\end{tabular}
\caption{Mesh properties (T6 finite elements) and times computing for the interface problem and various values of $\e$}
\end{center}
\end{table}

Even if only linear finite elements are necessary for the computations for the interphase problem, we can notice that
the CPU times necessary to obtain the solution quickly increases as the thickness of the interphase tends to 0.
The reason is that the mesh has to be sufficiently fine inside the interphase, at least four nodes along the thickness.
Therefore, in order to keep a reasonable condition number for the rigidity matrix, the mesh has to
be fine also in a large zone around the interphase. This necessity significantly increases the number of degrees of freedom
as the thickness of the interphase tends to zero.

For the interface problem, the computation is relatively independent of the parameter $\e$. As a consequence,
the meshes and the CPU times of the computations increase very slowly as the thickness tends to zero.
So, for small values of the thickness, the use of the imperfect interface model is very convenient.

\section{Numerical results for two elastic bodies glued}
 In  this section, we present examples of two elastic structures glued together. These examples
 are inspired by \cite{GRD08}.
 \medskip

\begin{figure}[!ht]
\centerline{\resizebox{6.cm}{!}{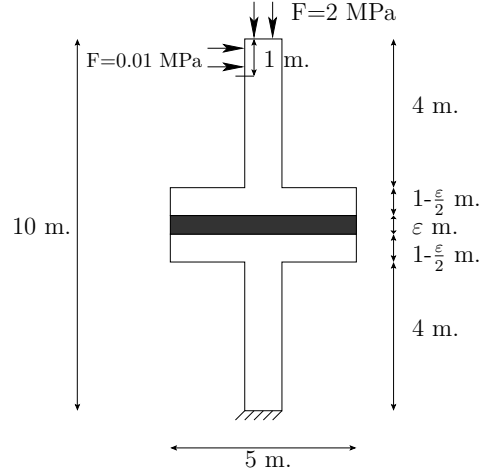}}
\caption{Exemple 2: geometry of the problem ($\e=0$ for the interface problem)}
\label{fig02}       
\end{figure}

 The first one is composed of two T-form elastic bodies (Aluminium) glued with an epoxy resin (see figure \ref{fig02} for
 the geometry). More precisely, the mechanical coefficients of the materials are as follows:

\begin{itemize}
\item In the adhesive (Epoxy resin): $\hat E= 4 GPa$, $\hat \nu=0.33$.

\item In the adherent (Aluminium): $E= 70 GPa$, $\nu=0.33$.
\end{itemize}

We present results for $\e=0.01$ m.

\begin{figure}[!ht]
\centering
\includegraphics[height=4.6cm]{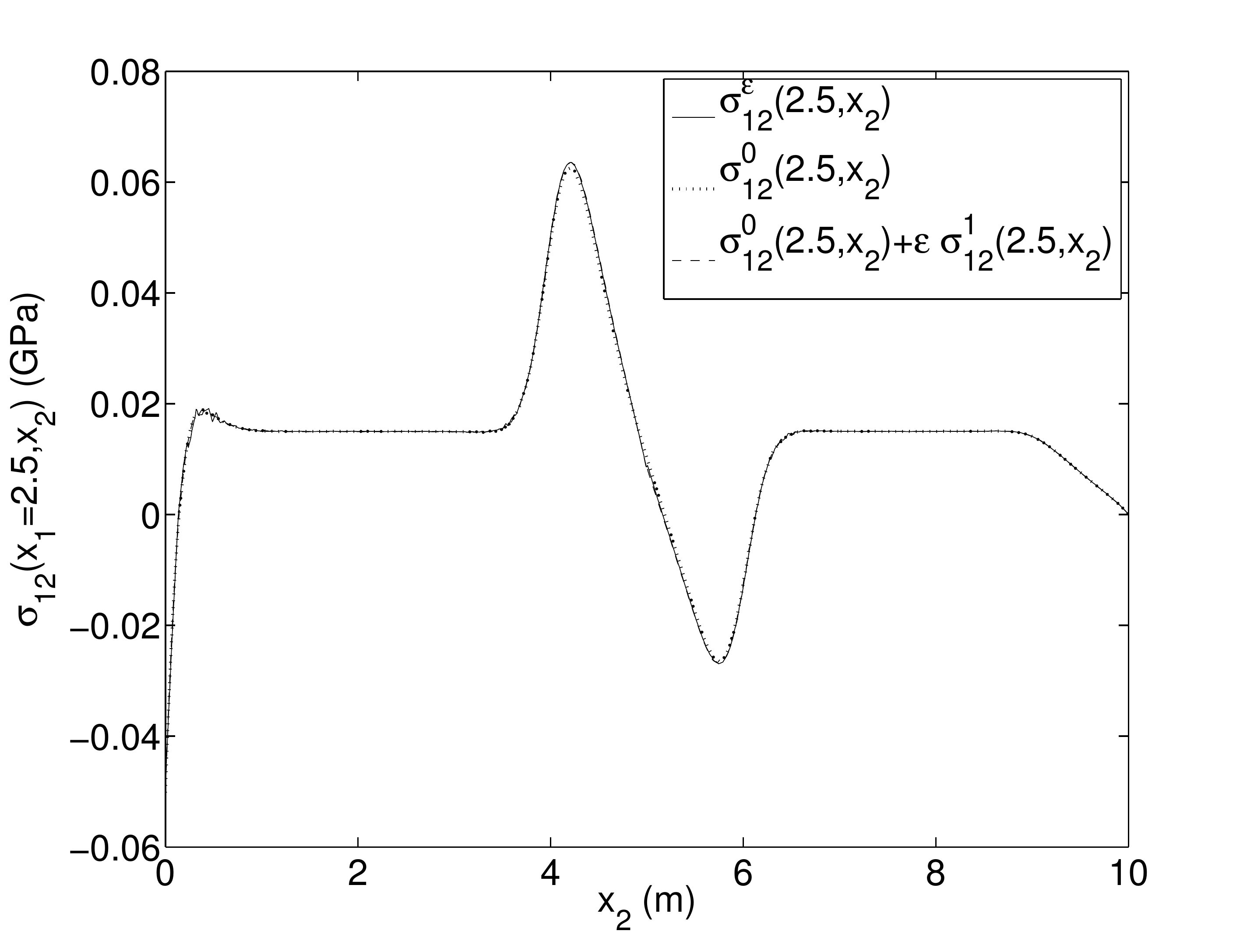}
\includegraphics[height=4.6cm]{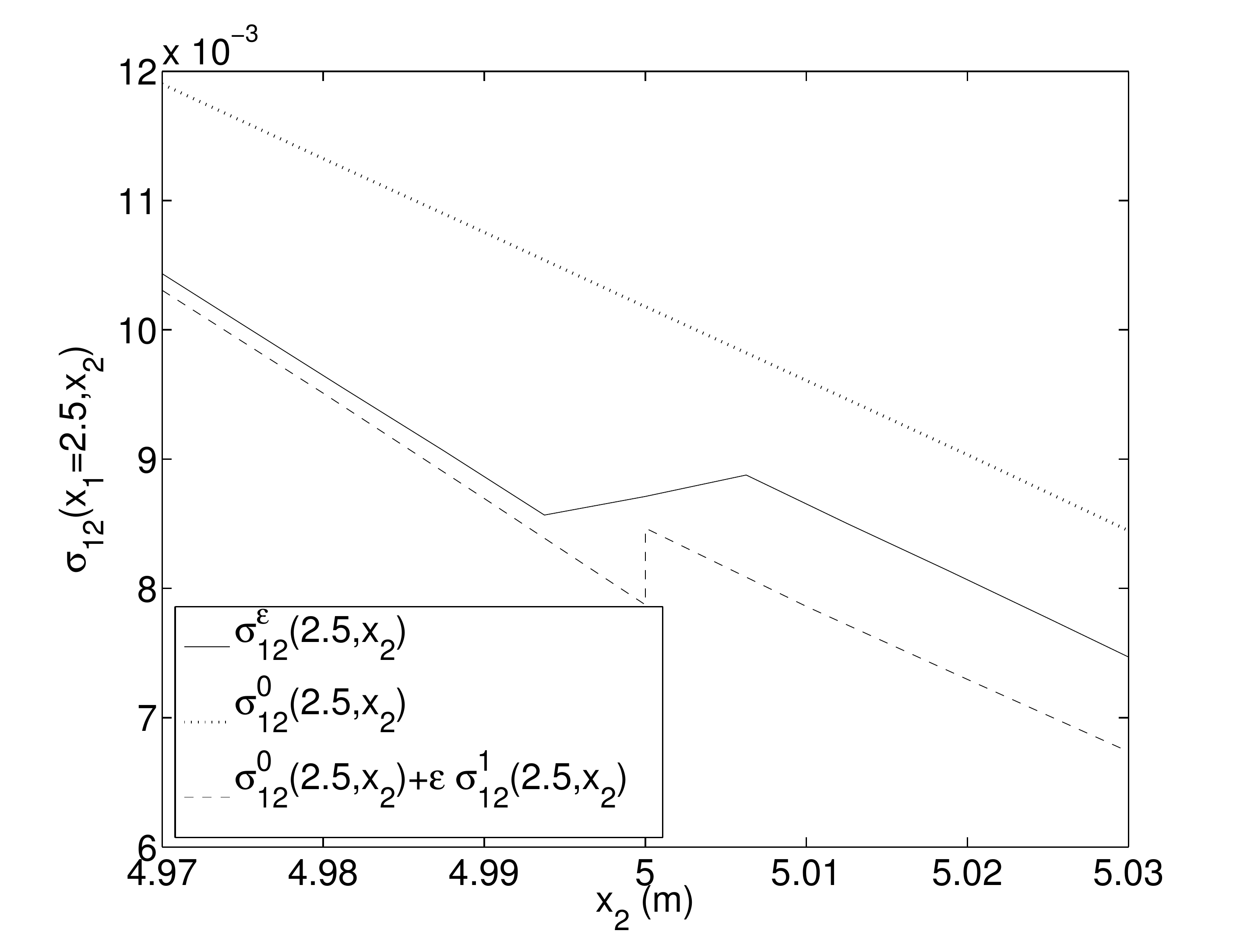}
\caption{Exemple 2 ($\varepsilon=0.01$ m) -- Stress $\sigma_{12}(x_1=2.5,x_2)$ (GPa) on a vertical cut }
\label{T1}       
\end{figure}

\begin{figure}[!ht]
\centering
\includegraphics[height=4.6cm]{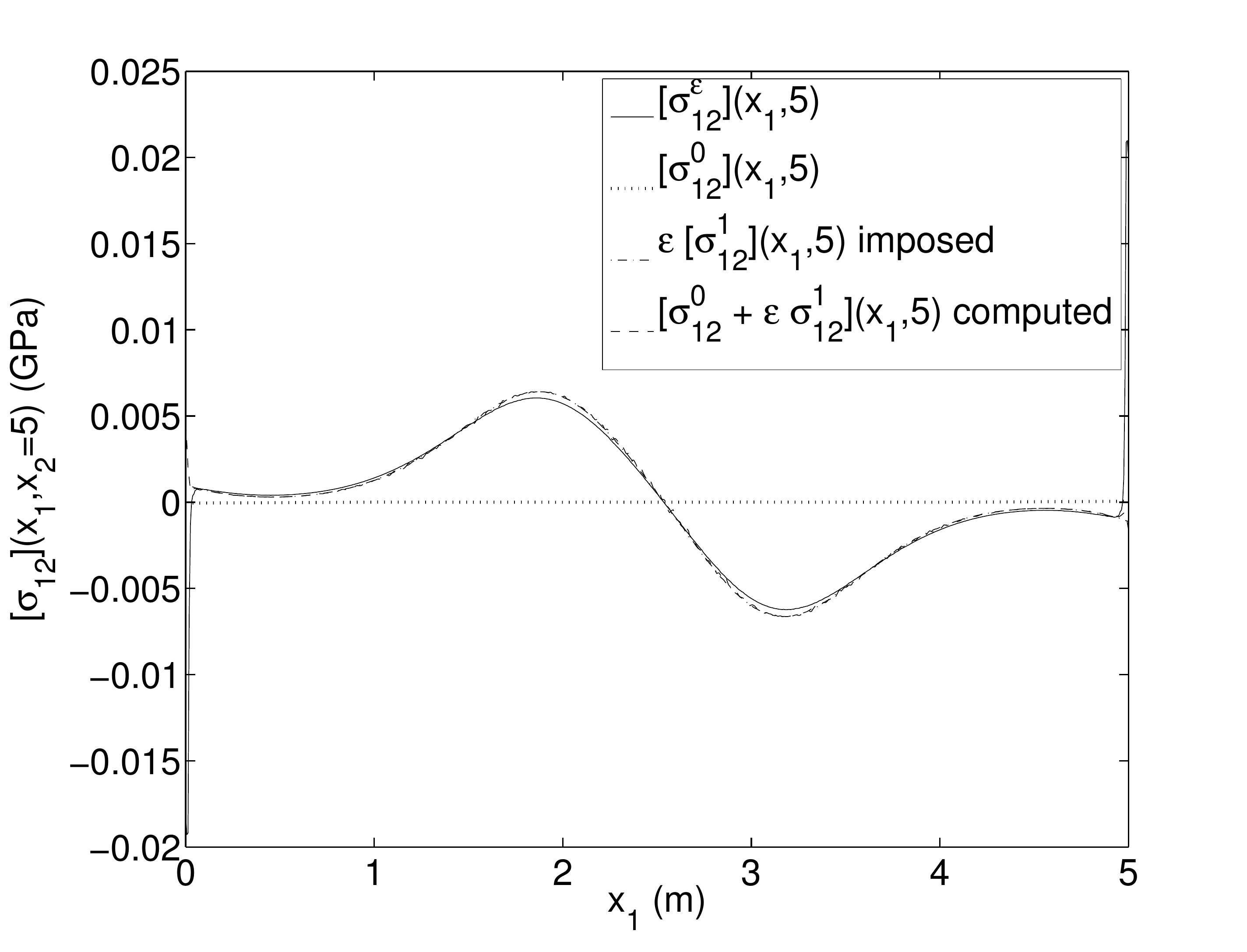}
\caption{Exemple 2 ($\varepsilon=0.01$ m) -- Jump in the stress $\lbrack \sigma_{12}\rbrack(x_1,x_2=5)$ (GPa) along the interface}
\label{T2}       
\end{figure}

\begin{figure}[!ht]
\centering
\includegraphics[height=4.6cm]{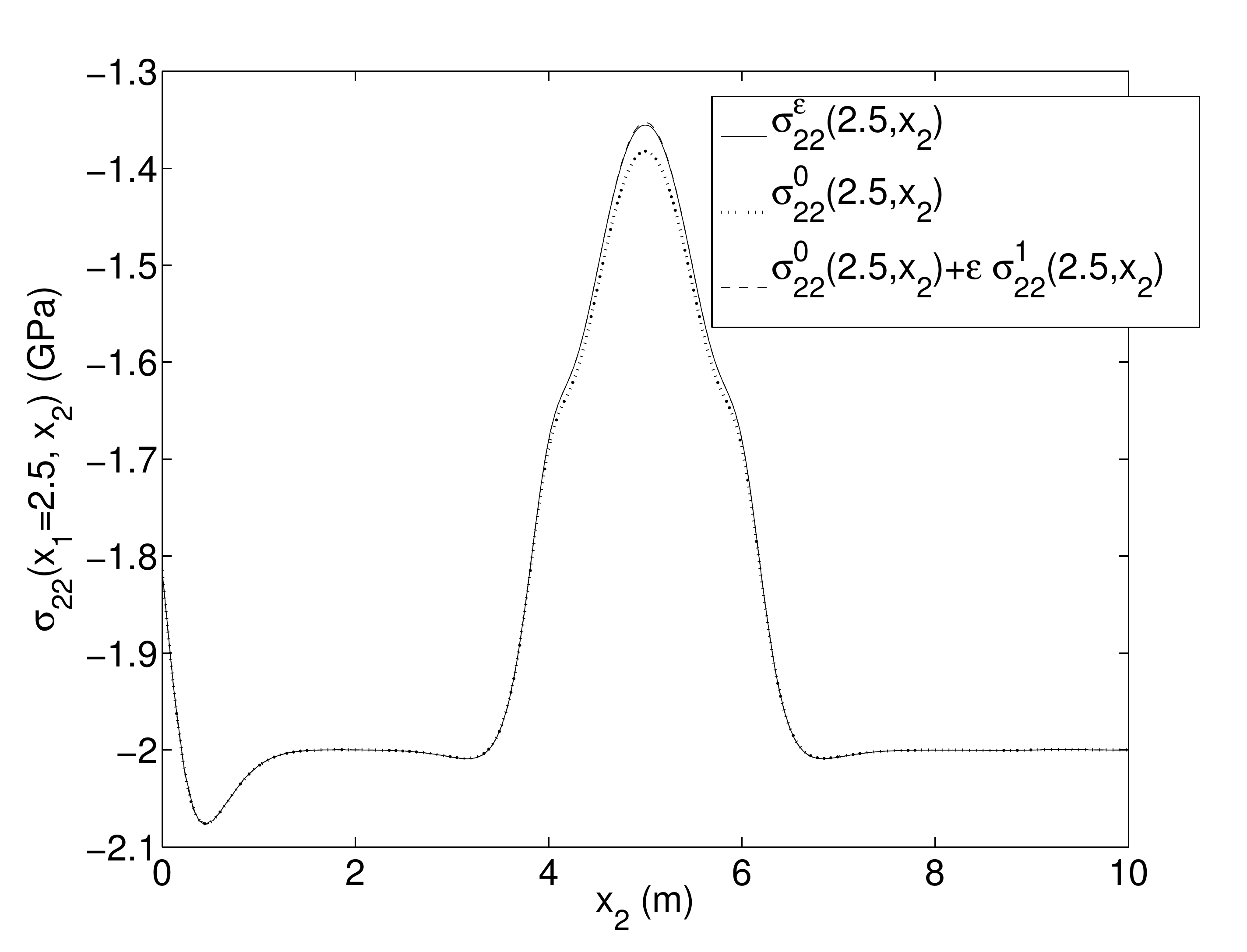}
\caption{Exemple 2 ($\varepsilon=0.01$ m) -- Stress $\sigma_{22}(x_1=2.5,x_2)$ (GPa) on a vertical cut}
\label{T3}       
\end{figure}

\begin{figure}[!ht]
\centering
\includegraphics[height=4.6cm]{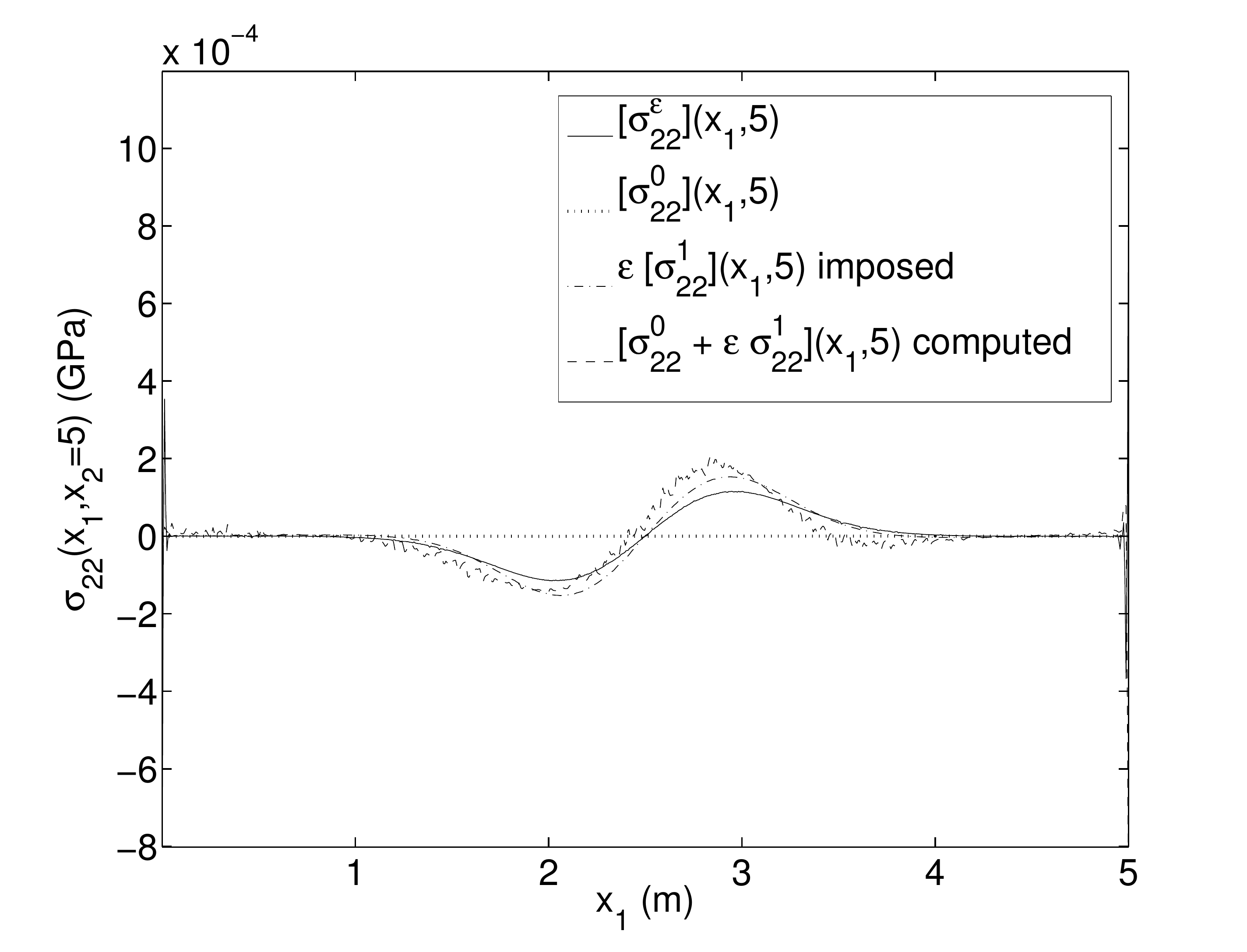}
\caption{Exemple 2 ($\varepsilon=0.01$ m) -- Jump in the stress $\lbrack \sigma_{22}\rbrack(x_1,x_2=5)$ (GPa) along the interface}
\label{T4}       
\end{figure}

\begin{figure}[!ht]
\centering
\includegraphics[height=4.6cm]{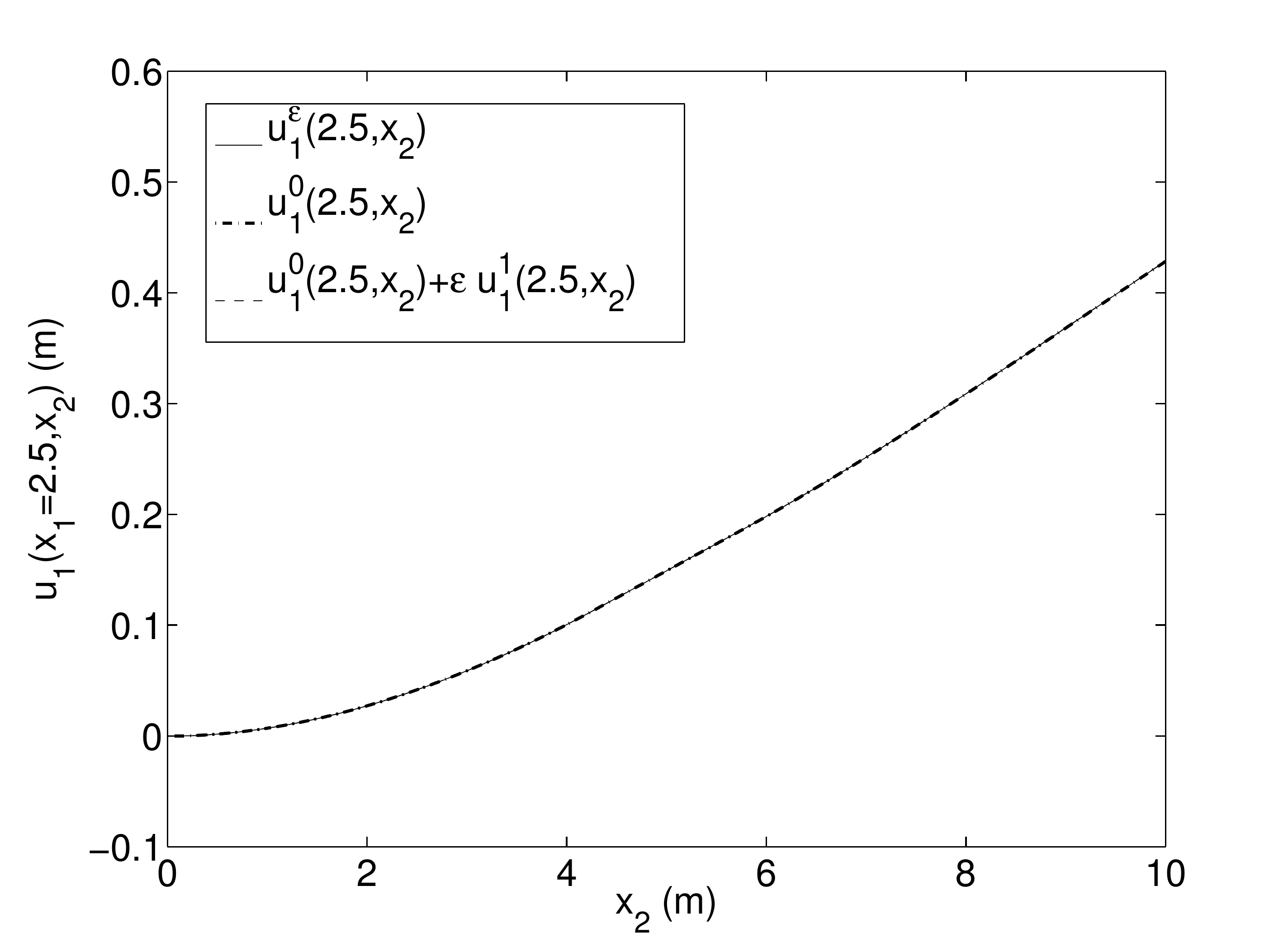}
\includegraphics[height=4.6cm]{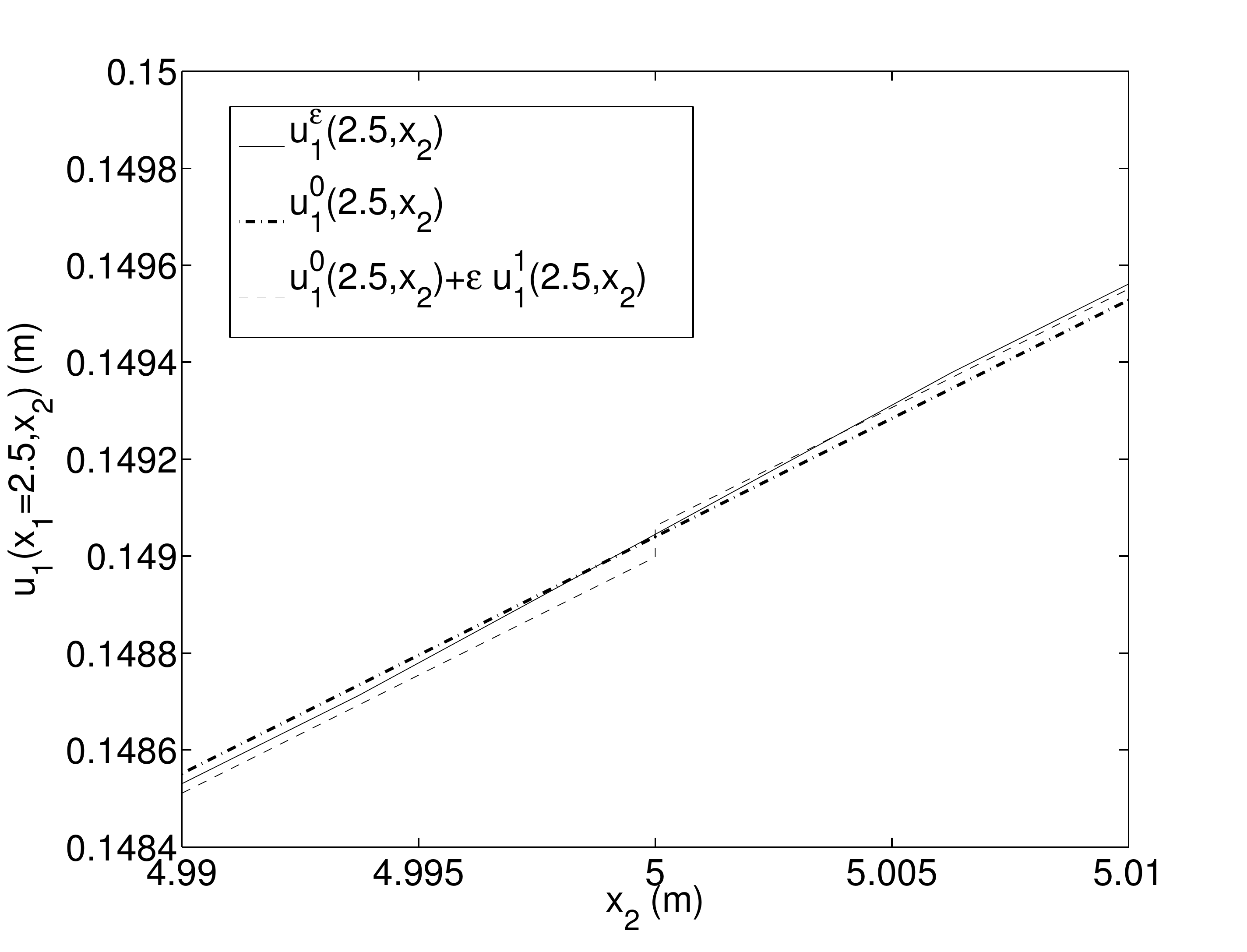}
\caption{Exemple 2 ($\varepsilon=0.01$ m) -- Displacement $u_{1}(x_1=2.5,x_2)$ (m) on a vertical cut (Zoom on the right)}
\label{T5}       
\end{figure}

\begin{figure}[!ht]
\centering
\includegraphics[height=4.6cm]{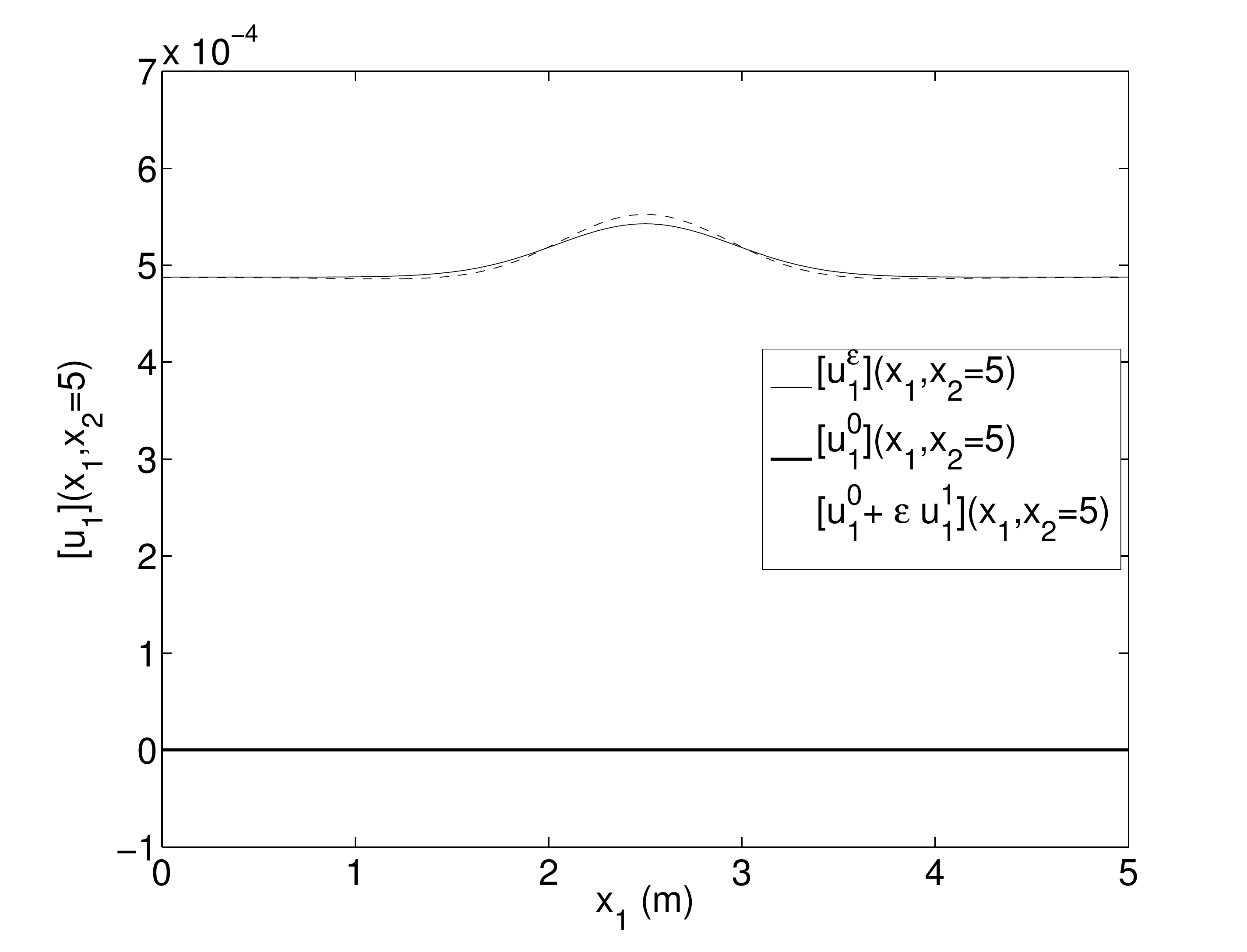}
\caption{Exemple 2 ($\varepsilon=0.01$ m) -- Jump in the displacement $\lbrack u_{1}\rbrack(x_1,x_2=5)$ (m) along the interface}
\label{T6}       
\end{figure}

\begin{figure}[!ht]
\centering
\includegraphics[height=4.6cm]{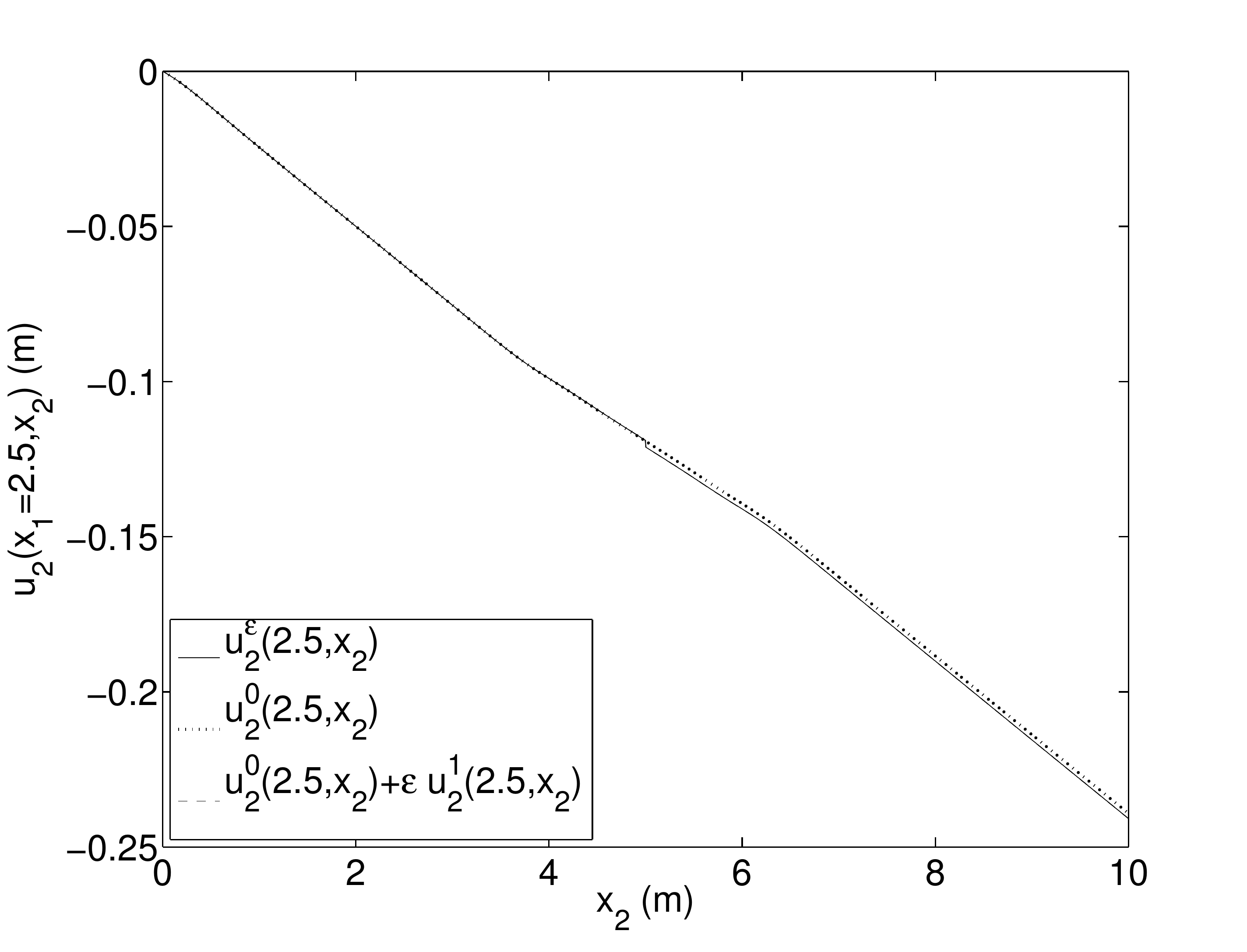}
\includegraphics[height=4.6cm]{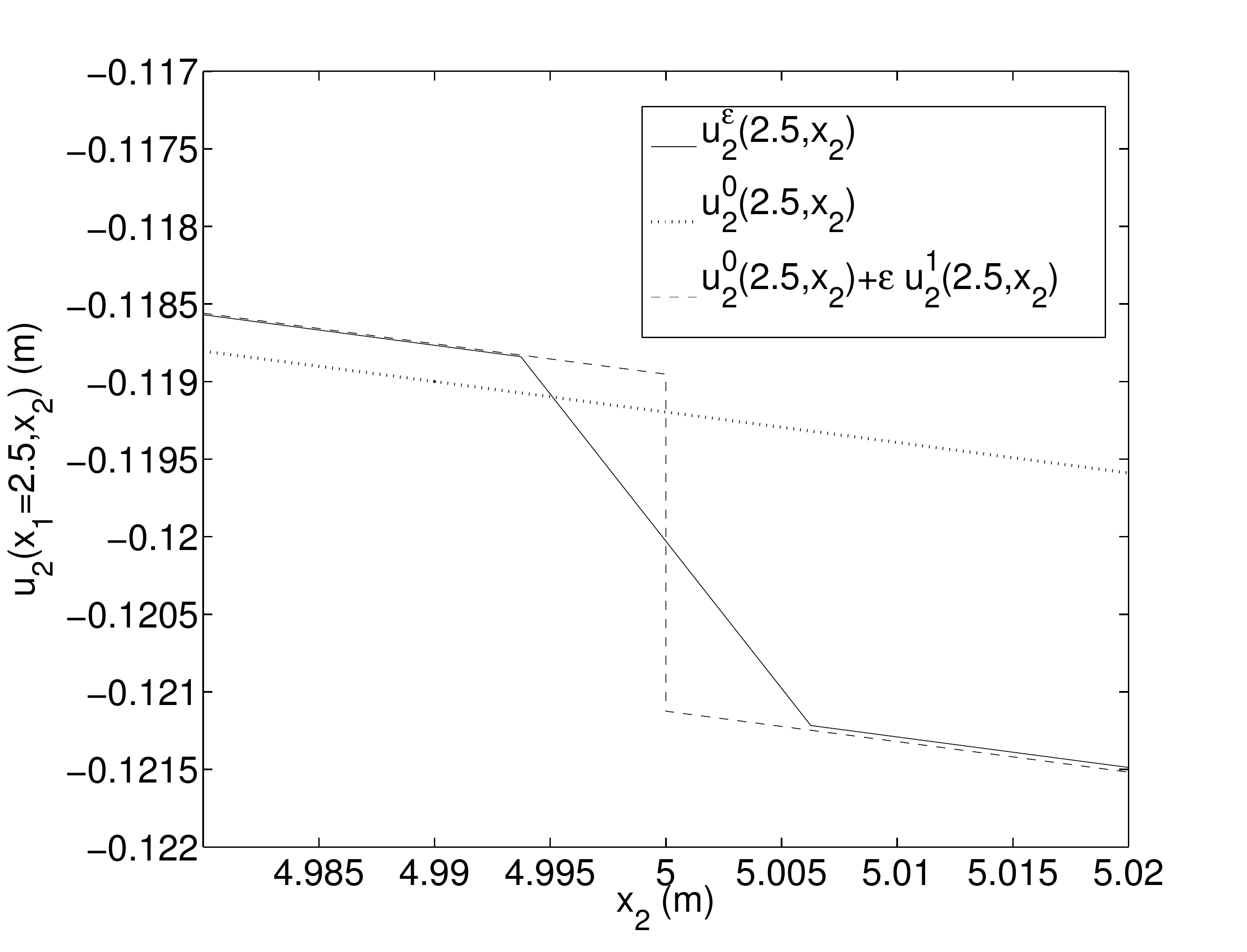}
\caption{Exemple 2 ($\varepsilon=0.01$ m) -- Displacement $u_{2}(x_1=2.5,x_2)$ (m) on a vertical cut (Zoom on the right)}
\label{T7}       
\end{figure}

\begin{figure}[!ht]
\centering
\includegraphics[height=4.6cm]{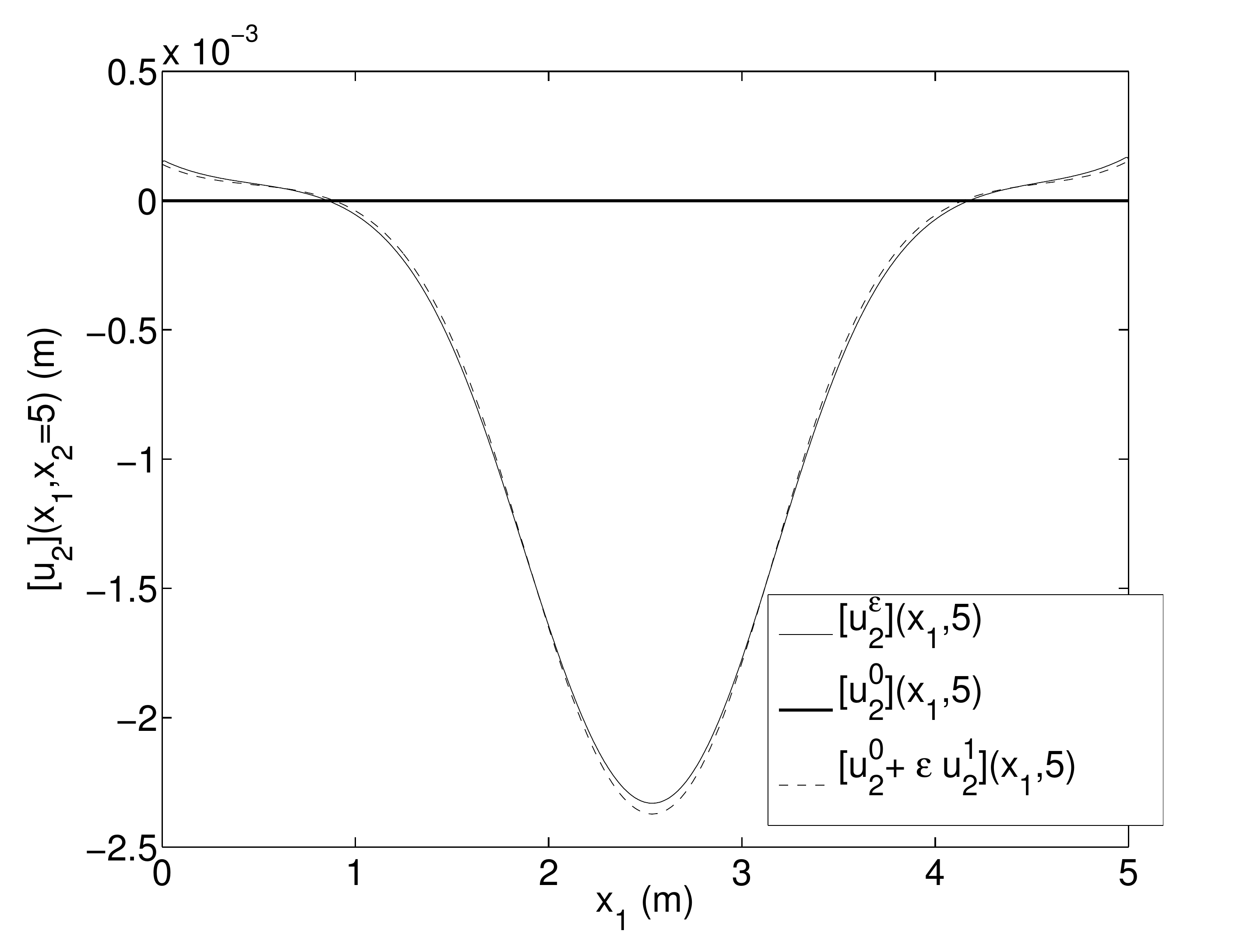}
\caption{Exemple 2 ($\varepsilon=0.01$ m) -- Jump in the displacement $\lbrack u_{2}\rbrack(x_1,x_2=5)$ (m) along the interface}
\label{T8}       
\end{figure}

The results show that the interface law at order 1 is able to reproduce the phenomena that occur in the interphase much more
accurately that the interface law at order 0.

For example, we observe that  $\sigma_{12}$ and the jump $\lbrack \sigma_{12}\rbrack$ are in good agreement with
 $\sigma^\e_{12}$ and $\lbrack \sigma^\e_{12}\rbrack$, respectively (see figures \ref{T1} and \ref{T2}).

 Even in the case of a small jump of a quantity (see figure \ref{T4} for $\lbrack \sigma_{22}\rbrack$, figure \ref{T6} for
 $\lbrack u_1\rbrack$), considering the order 1 term improve the results (see figure \ref{T3} for the stress $\sigma_{22}$ and figure \ref{T5}
 for displacement $u_1$).

 Finally, the model is able to reproduce the displacement jump $\lbrack u_2 \rbrack$ and the displacement $u_2$ with a very
 small error (see figure \ref{T7} and \ref{T8}).
\bigskip

In our last example, we consider the same geometry as before. The materials are now as follows:

\begin{itemize}
\item The adhesive is a reinforced Epoxy resin: $\hat E= 6 GPa$, $\hat \nu=0.33$.

\item The adherents are made of Polymethyl Methacrylate (PPMA) : ${E}= 3.1 GPa$, $\nu=0.4$.
\end{itemize}

In this last example, the elasticity coefficients of the components are very close. Let us observe that if they 
are identical, then the jumps conditions at order 0, that is to say $\lbrack u\rbrack=0$ and $\lbrack\sigma n\rbrack=0$, are exact conditions.
This example permit us to observe the improvement provided by the first order terms in the case of a small mechanical properties variation
between the glue and the adherents.

We present here only the results for the displacement $u_2$ (normal to the interface) and for the stress $\sigma_{22}$, because they are
representative of the results for the other components.

\begin{figure}[!ht]
\centering
\includegraphics[height=4.6cm]{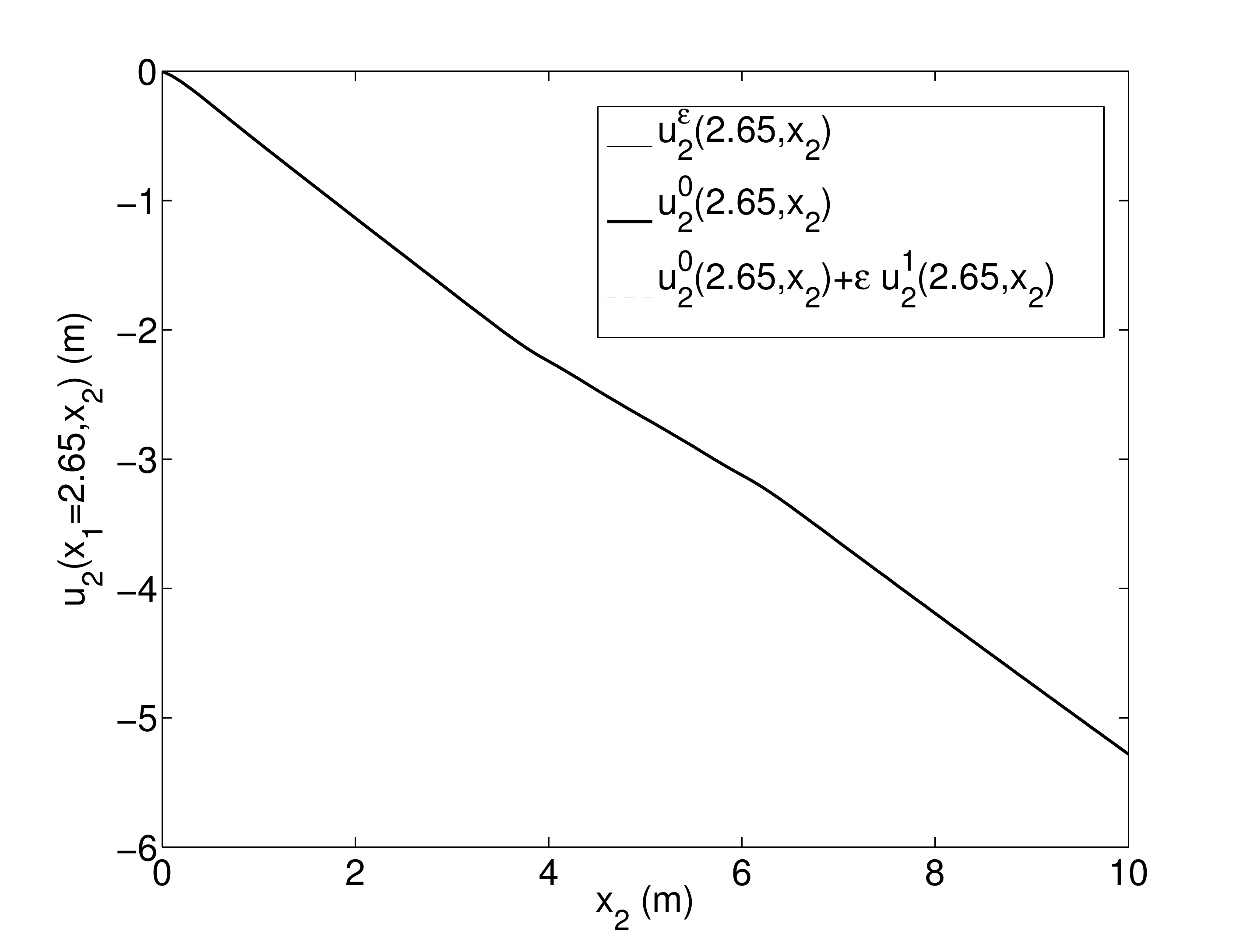}
\includegraphics[height=4.6cm]{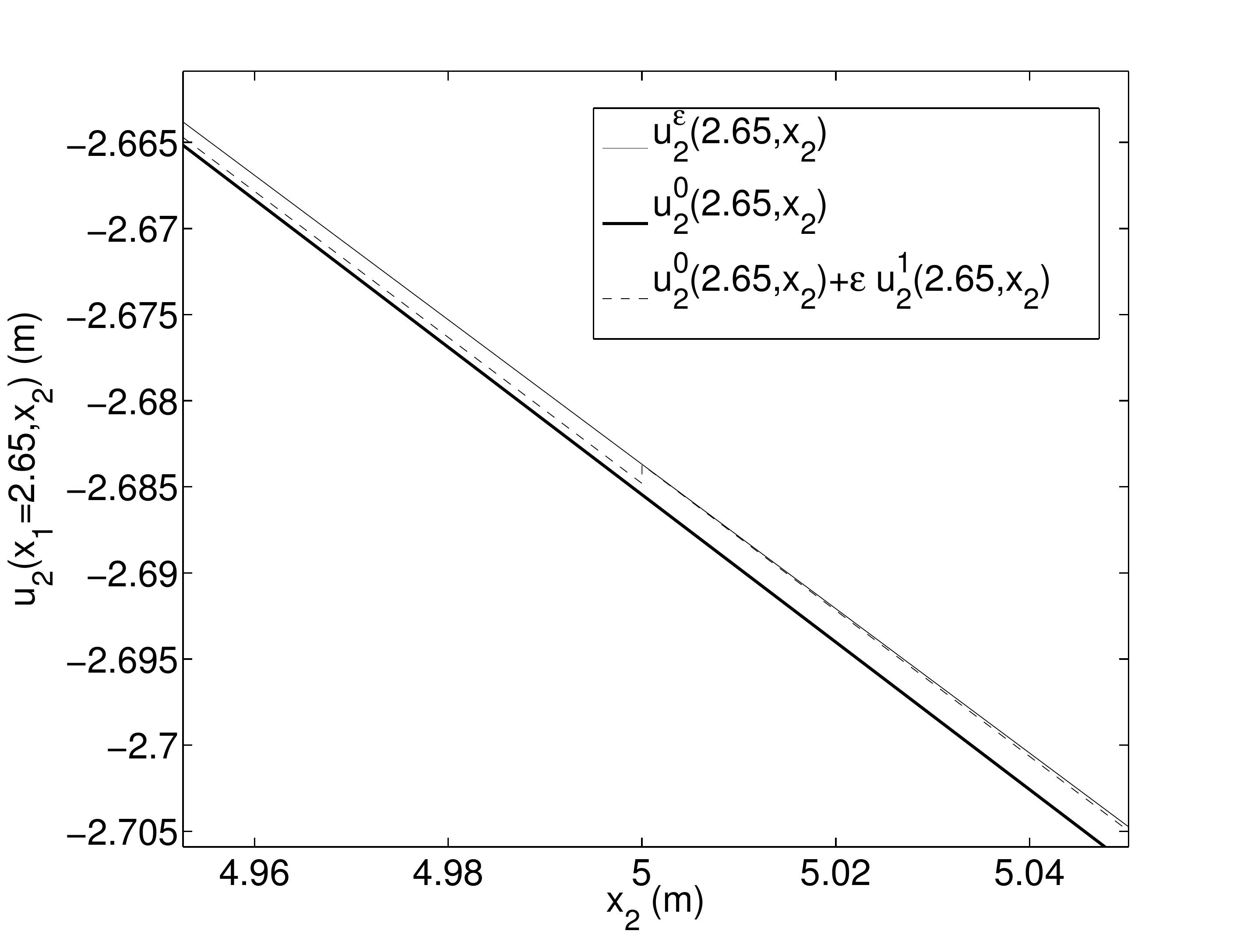}
\caption{Exemple 3 ($\varepsilon=0.01$ m) -- Displacement $u_{2}(x_1=2.65,x_2)$ (m) on a vertical cut (Zoom on the right)}
\label{T9}       
\end{figure}

\begin{figure}[!ht]
\centering
\includegraphics[height=4.6cm]{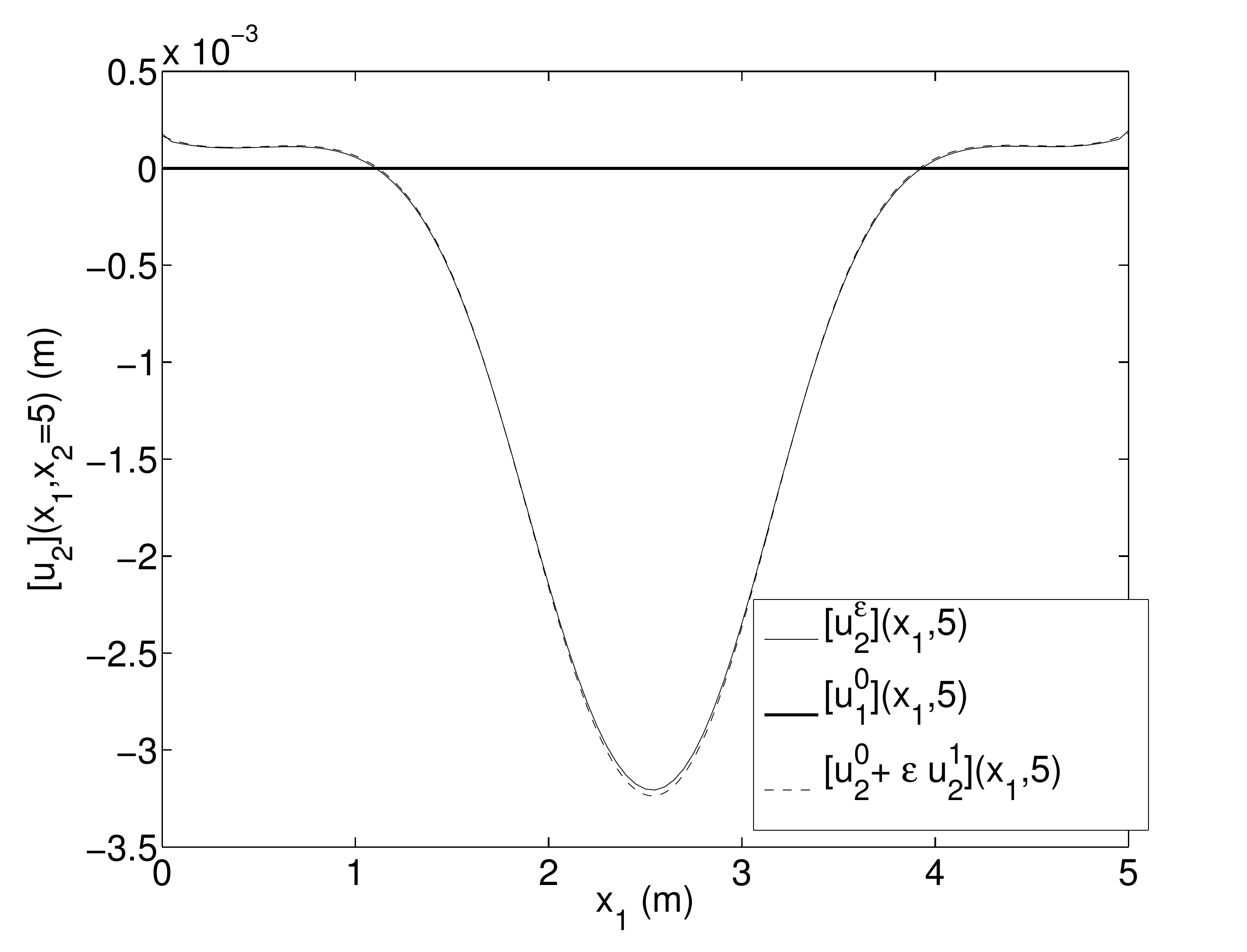}
\caption{Exemple 3 ($\varepsilon=0.01$ m) -- Jump in the displacement $\lbrack u_{2}\rbrack(x_1,x_2=5)$ (m) along the interface}
\label{T10}       
\end{figure}

\begin{figure}[!ht]
\centering
\includegraphics[height=4.6cm]{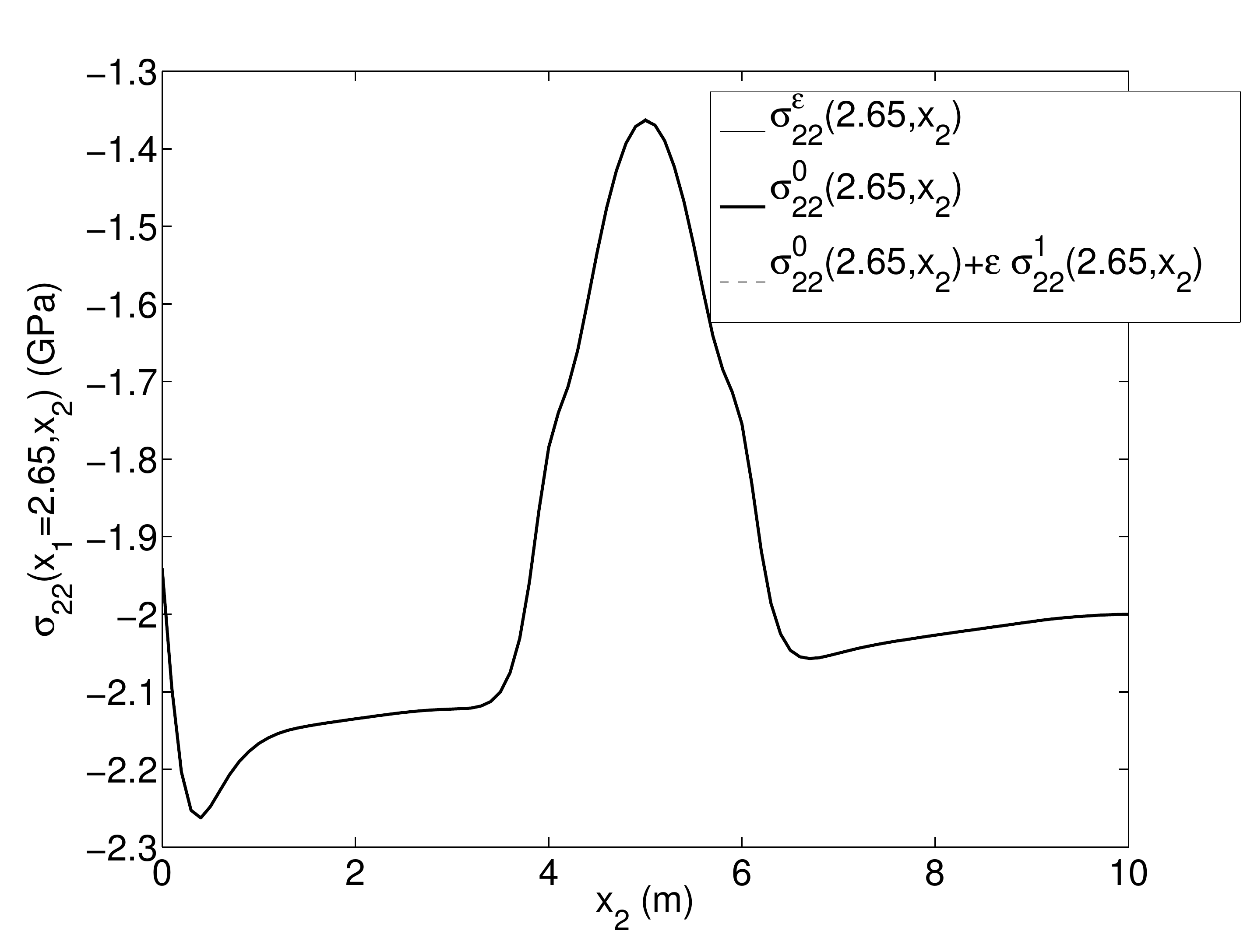}\hspace{0.2cm}
\includegraphics[height=4.6cm]{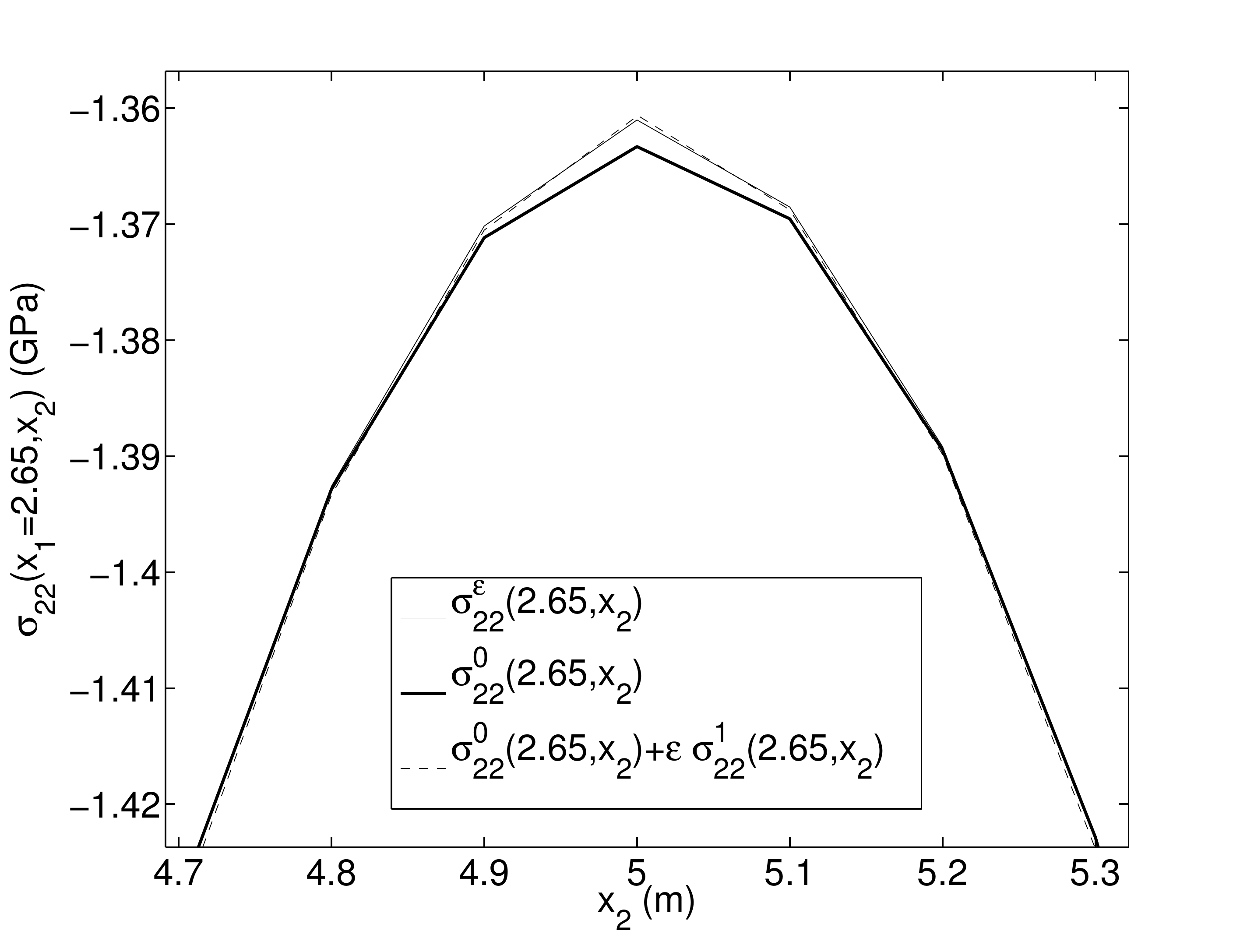}
\caption{Exemple 3 ($\varepsilon=0.01$ m) -- Stress $\sigma_{22}(x_1=2.65,x_2)$ (GPa) on a vertical cut (Zoom on the right)}
\label{T11}       
\end{figure}

\begin{figure}[!ht]
\centering
\includegraphics[height=4.6cm]{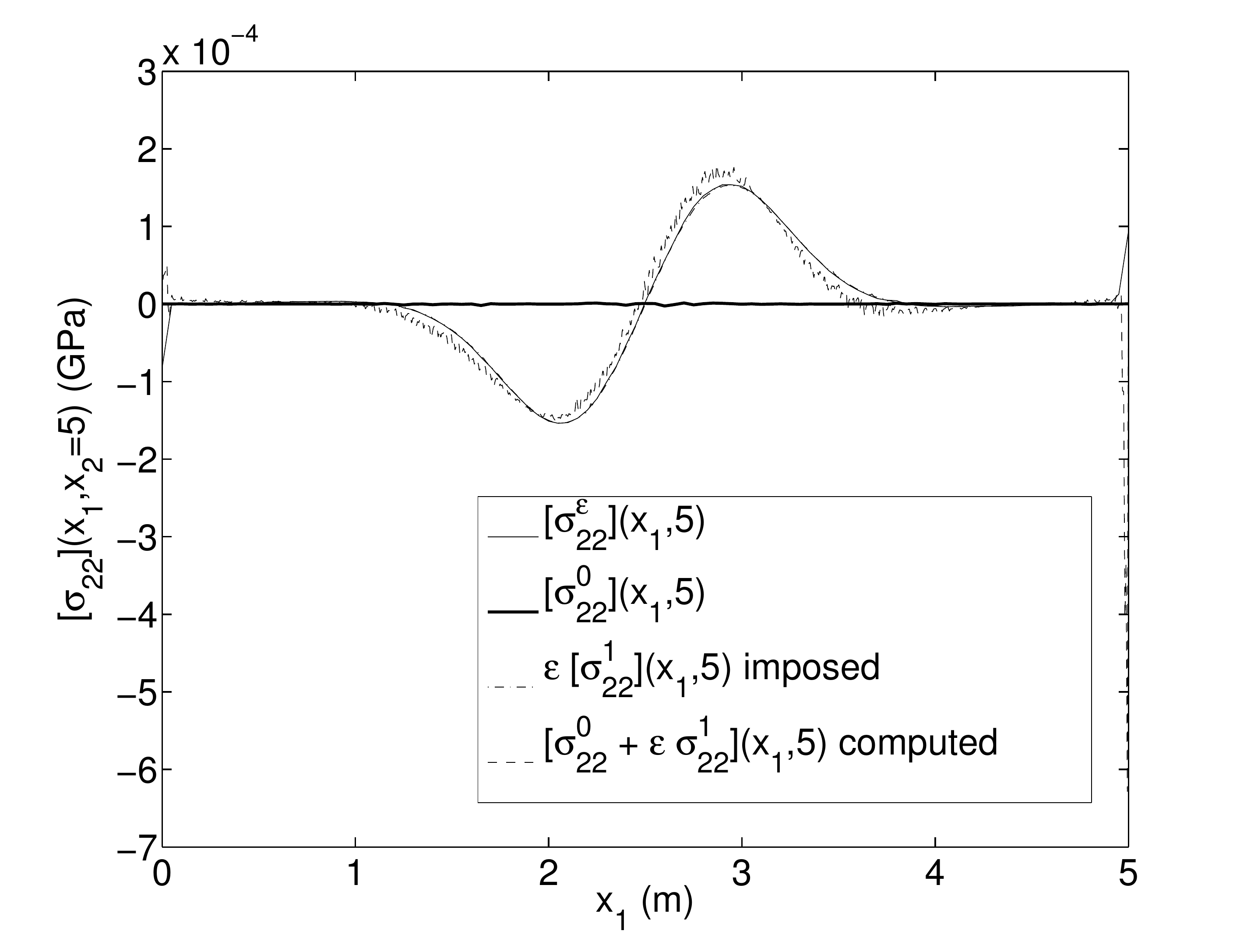}
\caption{Exemple 3 ($\varepsilon=0.01$ m) -- Jump in the stress $\lbrack \sigma_{22}\rbrack(x_1,x_2=5)$ (GPa) along the interface}
\label{T12}       
\end{figure}

%
%
%

Due to the small difference between the mechanical properties of the glue and the adherents, we can observe in figure \ref{T10}
that the jump in the displacement is very small compared to the size of the structure (less than $2\cdot 10^{-3}\%$ of the size of the structure),
and to the maximum of the displacement represented in figure \ref{T9} (less than $10^{-2}\%$ of the maximum of the displacement of the structure).
Nevertheless, considering the first order term improves the asymptotic approximation of the displacement (see figure \ref{T9}) and the stress
(see figure \ref{T11}). Moreover, the jump in the stress $\lbrack \sigma_{22}\rbrack$ is well approximated (see figure \ref{T12}). The oscillations that we can observe in figure \ref{T12}
are due to the smallness of the stress jump, that is of the same order of the error of the approximation.

\section{Conclusion}

In this paper, we have presented an interface law at order 1 when the Lamé's coefficients of the adhesive do not rescale with the thickness
of the interphase.  Based on the proposed interface law, some
 numerical experiments were also presented that show the accuracy of the method
when the interphase thickness becomes smaller and smaller.
For this purpose, we have developed an original numerical scheme based on the Nitsche's method to simulate the adhesion between two elastic materials.

This model is very efficient (maximum relative error less than 1 \% for a ratio of thickness smaller than 1 \%) and the interface law is able to reproduce the mechanical behavior of the real interface.
On the other hand, the numerical model developed in this paper is less expensive than the solution of the real problem.
More precisely, the method is independent of the thickness of the interphase and it becomes more and more efficient as the thickness decreases.
For example, in the first numerical test proposed above, the CPU times of the interphase problem and the asymptotic interface problem are equivalent when $\e=0.1$, but their ratio is lower than $\frac{1}{10}$ when $\e=0.001$.







\section*{Acknowledgement}
RR  thanks  the financial support of the Italian Ministry of Education, University and Research
(MIUR) through the PRIN project “Multi-scale modeling of
materials and structures” (code 2009XWLFKW).

\section*{References}
\bibliographystyle{elsarticle-harv}
\bibliography{DLR}

\end{document}